\documentclass[a4paper]{article}
\usepackage{graphicx}
\usepackage{amsmath, amssymb}
\usepackage{amsbsy}
\usepackage{xcolor}
\usepackage{siunitx}
\usepackage{csquotes}
\usepackage{authblk}
\usepackage{a4wide}

\newcommand\Gam{\mbox{$\it \Gamma$}} 

\title{Thermal boundary layers in turbulent Rayleigh-Bénard convection with rough surfaces}

\author[1]{R.~du~Puits}
\author[1]{A.~Loesch}
\author[2]{J.~Salort}
\author[2]{F. Chilla}
\affil[1]{Technische Universitaet Ilmenau, Institute of Thermodynamics and Fluid Mechanics, P.O. Box 100565, 98684 Ilmenau}
\affil[2]{Laboratoire de physique, UMR CNRS 5672, École Normale Supérieure de Lyon, 46 allée d'Italie, 69364 Lyon cedex 7, France}
\date{}                     
\date{\today}

\begin{document}

\maketitle

\begin{abstract}
We present highly resolved measurements of the near-wall temperature field in thermally driven convection at a rough surface. Our measurements have been undertaken in a very large experimental facility called the ``Barrel of Ilmenau''. They provide a unique insight into the local transport process at the interface between a hot solid surface and a surrounding fluid. In order to probe the near-wall temperature field, we used a tiny micro-thermistor of 130~µm in diameter and 330~µm in length with a response time $\tau_{70}$ of less than 150~ms. This sensor is forty times smaller than the thickness of the boundary layer, and it permits a resolution better than the typical Kolmogorov micro-scales that appear in our experiment. We demonstrate that the heat flux enhancement generally observed at rough surfaces, basically results from an increase of the local heat transfer coefficient at the top of the roughness elements. We also shed light on the detailed physical process behind the transition in the scaling of the global heat transfer relation $Nu\sim Ra^{\gamma}$ that has been observed beyond a critical Ra number $Ra_c$. As reported by [Tisserand {\it et~al.} {\em Phys. Fluids \/} 23, 015105(2011)], it already appears at $Ra_c\approx 10^{10}$. We found this transition at rough surfaces to be attributed to two phenomena, i) an earlier laminar-turbulent transition of the boundary layer at the top of the roughness elements, and ii) a modification of the temperature field in between the obstacles.

\end{abstract}

\section{Introduction}

Thermal convection at a rough surface is a paradigm for a great variety of heat transfer processes in natural and engineering environment. As well as in the case of a smooth surface, the temperature field in the close vicinity of the wall is the key to understand and to predict the magnitude of the convective heat flux from a solid body to a surrounding fluid. Although convection at rough surfaces has been studied quite frequently in the past, experimental data being highly resolved in space and time are rare and frequently obtained at very specific or badly reproducible applications. Our work provides a set of temperature data obtained in confined natural convection that meets such requirements. The particular aim of this work is to contribute to a more universal understanding of such heat transfer processes and to improve their general predictability.

The heat transfer in many geophysical and engineering applications occurs at surfaces that are not smooth. For instance, the heat exchange between an urbanized or tree covered landscape to the atmosphere represent such a process. Nowadays, there is not yet a simplified and universal model predicting the heat flux precisely. Since this quantity is one of the main contributions to the total energy balance of the Earth, it directly affects the mean temperature in the atmosphere. Hence, uncertainties in the prediction of the convective heat transfer rate bias the prognosis of the global warming. Heat exchangers, in which rough surfaces increase the efficiency of the heat transfer represent another typical example with no less economical and ecological importance. For instance, they are widely used in huge facilities to generate electrical energy, but they are also installed in billions of households to produce heat using gas and oil burners. The literature on this subject is as diverse as the specific applications of thermal convection at rough surfaces. It is not our intension here, and probably it is even impossible, to give a comprehensive review on the literature in this field. Instead, we will refer to a few typical examples from Geophysics and Engineering as well as to a few fundamental studies that show the great diversity of studies done in the past.

For instance, there are many publications in geophysical journals that address the convective heat transfer between the Earth's ground and the atmosphere. One of the first studies of this subject has been carried out by Thom {\it et al.} \cite{Thom1975}, who investigated the fluxes of sensible and latent heat over a level forested region. The authors applied the so-called aerodynamic method of flux determination (a profile-gradient based model), but they had to conclude ``that the complexity and uncertainty of obtaining reliable flux estimates over tall vegetation by the aerodynamic model preclude its use ...''. About one and a half decades later, Schumann proposed a simple and more general model to describe a convective boundary layer above a homogeneously rough ground at zero mean wind \cite{Schumann1988}. His model was derived from momentum and heat balances in the atmospheric surface layer and incorporates closures based on Monin-Obukhov theory. It provides explicit relations of the turbulent temperature as well as the velocity fluctuations with respect to the vertical distance from the ground. Moreover, it proposes scaling laws $Nu\sim Ra^{\gamma}$ with $\gamma$ being $1/3$ for a smooth and $1/2$ for a rough surface. In their paper, the authors also suggest a limit $Ra_t$ to discriminate between smooth and rough surfaces:
\begin{equation}
Ra_t>560\left(\frac{H}{z_0}\right)^{8/3}, \hspace{1cm} \textrm {for~ Pr}=0.7.
\label{limitsmallrough}
\end{equation}
The Rayleigh, the Nusselt and the Prandtl number are defined as:
\begin{equation}
Ra=\frac{\beta g\Delta \vartheta H^3}{\nu\kappa}, \hspace{1.0cm} Nu=\frac{\dot{q}_{k}}{\dot{q}_{d}}, \hspace{1.0cm} Pr=\frac{\nu}{\kappa} .
\label{Definition_Ra_Nu}
\end{equation}
In these definitions, the variable $H$ is the interfacial layer of the atmosphere enclosing the layer with the temperature inversion (or in a more general view the total thickness of the fluid layer), $z_0$ is the height of the surface roughness, $\beta$ is the thermal expansion coefficient, $g$ is the gravitational acceleration, $\Delta \vartheta$ is the temperature drop across the fluid layer and  $\nu$ and $\kappa$ are the kinematic viscosity and the thermal diffusivity of the fluid, respectively. The Nusselt number is defined as the ratio between the convective and the diffusive heat fluxes $\dot{q}_{k}$ and $\dot{q}_{d}$, the Prandtl number is the ratio between the kinematic viscosity and the thermal diffusivity. In a more recent work, a group of researchers calculated numerical simulations to investigate turbulent convection at periodic as well as randomly structured rough surfaces \cite{Toppaladoddi2014}. The primary aim of this work was to understand the effect of various kinds of roughness on the magnitude of the turbulent fluxes of heat and momentum and how the shape of roughness may change the melting or the growth of the Arctic see ice in the oceans. The authors developed a two-dimensional MPI code that is based on the Lattice Boltzmann method, and they applied this code to channel flow and turbulent Rayleigh-B\'enard convection. For the latter problem they found a variation of both, the scaling exponent $\gamma$ as well as the pre-factor $C$ in the $Nu=C~Ra^{\gamma}$ scaling law and they identified an enhanced plume production from the tips of the roughness elements to be the reason for this variation.

In the engineering community, the diversity of convective heat transfer problems at rough surfaces is even broader. It covers convection at the outer side of buildings, the heat transfer in heat exchangers or the cooling of turbines and power electronic circuits, just to mention a few examples. In a recent work published in Applied Thermal Engineering, Palyvos investigated the thermal losses from building surface or roof mounted solar convector to the ambient air due to wind induced convection \cite{Palyvos2008}. In this reviewing paper, the author critically discusses various existing correlations to estimate the convective heat transfer coefficient of buildings, and he proposed an own, alternative model. He also recognized that ``almost all these correlations are based on empirical data and exhibit a significant lack of physical generality''. Consequently, Palyvos states that ``More realism is needed, i.e. field rather than laboratory measurements, as well as some sort of standardization in the choice of such things as the wind velocity sensors or the measurement topology, e.g. height above ground and/or distance from the fa\c{c}ade wall or roof, etc. In this way, a much smaller set of well proven and generally accepted correlations may emerge, which will greatly help the designer/modeler.'' Following the request towards more universality, Ghodake investigated the enhancement of the convective heat transfer coefficient in heat exchangers with various ribs numerically \cite{Ghodake2016}. Using ANSYS CFD, the authors analysed different shapes of ribs such as V-shaped, triangular and rectangular as well as semicircular ones. They also varied the angle of attack of the flow. The most general conclusion of this, rather empirical case study, is the hypothesis that the enhancement of the heat transfer is associated with an increase of separations and reattachments over the ribbed wall. This boosts the fluid mixing, creates flow unsteadiness and interrupts the development of the thermal boundary layer. Another study of the convective heat transfer at rough surfaces which was dedicated to flat and finned heat sinks for the cooling of semiconductors has been reported by a group of Italian researchers \cite{Ventola2014}. Unlike in the references above, the authors investigated a grainy surface as to be formed from a direct metal laser sintering process. Using a self-made test facility, they measured the convective heat transfer coefficient at a great variety of surfaces with different grain. In particular, they deserve credit for developing a theoretical model that is based on a surface fully covered by sand of uniform grain (as used in \cite{Schlichting2000}. The key idea of this general model is ``to estimate the size of the eddies that dominate the heat transfer close to the wall by a combination of the size of the roughness elements, i.~e. $k_p$ and $k_a$ (the peak surface roughness and the grain size diameter),
and the viscous length scale $\eta$. In the conclusion of their work, the authors state that the observed enhancement of the heat transfer at grainy surfaces could not be explained by the increase of the effective surface area but must be caused by a variation of the fluid-mechanical properties of the boundary layer.

While a great number of papers deals with specific applications of thermal convection at rough surfaces, only a few publications focus on this problem from a rather fundamental perspective. In the late 1960s, Townes \& Sabersky  and Gowen \& Smith conducted laboratory experiments, in which they investigated canonical model flows over rough surfaces \cite{Towns1966,Gowen1968}. They measured the temperature and the velocity field in the very thin convective boundary layer. But, the experiments were insufficient to discriminate, e.~g. between diffusive and turbulent transport because of the lack of an adequate spatial and temporal resolution. In the last two decades it became also very popular to study natural convection with rough surfaces in the so-called Rayleigh-B\'enard(RB) set-up. This system consists of a fluid layer of the thickness $H$ that is heated from below and cooled from the top. Convection sets in due to the density difference between the hot fluid at the bottom and the cold fluid at the top plate. Only two control parameters exist to describe the model flow completely, the Rayleigh number $Ra$ and the Prandtl number $Pr$ (for the definitions see Eq.~\ref{Definition_Ra_Nu}). In many applications as well as in all laboratory experiments, the fluid layer is laterally confined by sidewalls. In this case, the aspect ratio $\Gam$, being the ratio between the lateral extent of the fluid layer $D$ and its thickness $H$ completes the set of control parameters. The first researchers who introduced roughness to the Rayleigh-B\'enard set-up were groups in Hongkong \cite{Shen1995,Du2000} and Lyon \cite{Ciliberto1999}. Shen et al. used a pair of rough plates structured by a lattice of square pyramids with a fixed height $h_s$ and a fixed spacing $d_s=2h_s$ at their surface and build up RB experiments of aspect ratios $\Gam=1$ and $\Gam=0.5$. They measured the near-wall temperature and velocity field in these two test cells applying a special particle tracking technology that is based on thermo-sensitive particles. In particular, they measured the temperature of the fluid as a function of the vertical distance $z$ from the horizontal plates using a small movable thermistor (Thermometrics, AB6E3-B10KA202J). The essential result of their work is the discovery that the emission of  plumes completely changes compared against a smooth surface. Furthermore, they found that the horizontal shear flow along the rough surface creates eddies in between the pyramids, which enhances the detachment of thermal plumes and increase the total heat transport with respect to the smooth surface. Furthermore, they found that this process does not change the exponent $\gamma=2/7$ in the $Nu\sim Ra^{\gamma}$ scaling law. Ciliberto and Laroche investigated spherical roughness elements randomly or periodically arranged at the surface of the heated bottom plate. Unlike Shen et al., they reported an increase of the exponent to $\gamma>2/7$ for randomly distributed obstacles. We wish to mention here three more references on turbulent RB convection with roughness showing the full discrepancy that actually exists on this field.  Qiu {\it et al.} and Wei {\it et al.} reported an increase of both, the exponent $\gamma$ as well as the pre-factor $C$ of the $Nu=C Ra^{\gamma}$ relation  \cite{Qiu2005,Wei2014}.  Roche {\it et al.}  even observed an exponent $\gamma=0.5$ in a RB cell with 110~µm deep V-shaped grooves at the top and bottom plates \cite{Roche2001}. The cell was operated with gaseous Helium at a temperature of about  5~K and the transition in the exponent starts at a Ra number of $Ra\approx10^{12}$. The authors explain the change in the scaling exponent with a laminar--turbulent transition of the boundary layer. However, they could not verify this hypothesis, since direct measurements of the near-wall flow field and, hence, the proof of the laminar-turbulent transition were not feasible in this experiment. Another kind of roughness elements that are also used in our own experiment, has been introduced by Tisserand {\it et al.} \cite{Tisserand2011}. The researchers operated a water-filled RB cell at room temperature and they prepared only the surface of the heated bottom plate with cubic-shaped elements of a height corresponding to the thickness of the thermal boundary layer. The cuboids were arranged in a square array of square plots with a period of twice the length (width) of the elements. The particular idea of this, asymmetric cell was to compare the thermal impedances of the lower half of the cell (rough) and the upper half of the cell (smooth), when they interact with the same bulk flow. In summary of their measurements, the authors conclude that the convective heat transport between a solid surface and a fluid is largely a local process and does virtually not depend on the global structure of the flow. They also could confirm the observation from all previous experiments that the enhancement of the Nusselt number exceeds the increase of the surface area due to the roughness. Salort {\it et al.} and Liot {\it et al.} continued the work on this specific roughness structure \cite{Salort2014,Liot2016}. In a collaborative work between the \'Ecole Normale Sup\'erieure de Lyon and the Technische Universitaet Ilmenau, the researchers measured the global heat transport as well as the local temperature and the velocity fields in the vicinity of the rough surface. The measurements of the global heat flux and a first, however poorly resolved measurement of the near-wall temperature field has been undertaken in the Lyon water cell. The local velocity field has been measured in the large-scale RB experiment called the ``Barrel of Ilmenau''. Because of its large size, the diameter amounts to 7.15~m and the height amounts up to 6.30~m, it actually provides the best spatial resolution of any kind of boundary layer measurement in thermal convection at very high Ra numbers. The aim of this joint work was to understand the specific mechanism, how does the convective heat transfer at the rough surface get enhanced by the destabilization of the boundary layer. In fact, the authors could validate that because of the roughness elements the critical Ra number $Ra_c$ decreases, and the boundary layer becomes turbulent. They also found a significant heat transfer enhancement beyond this bound, although the scaling exponent $\gamma$ does not increase to the asymptotic limit $\gamma_{asymp}=0.5$ as predicted by Goluskin \& Doering in their upper bound analysis \cite{Goluskin2016}. Another systematic investigation of the effects of roughness geometry on turbulent RB convection over rough plates has been reported by Xie \& Xia very recently \cite{Xie2017}. The authors applied pyramid-shaped roughness elements periodically distributed at both plates. Their measurements cover Ra numbers between $7.5 \times 10^7<Ra<1.3 \times 10^{11}$ and Prandtl numbers between $3.57<Pr<23.34$. The authors of this paper classified the heat transport scaling into three regimes. Regime~1 is considered to be the dynamically smooth regime with a $Nu\sim Ra^{\gamma}$ scaling equally to the smooth case. In the Regimes~2 and 3, the scaling of the heat transport is enhanced and the width-to-height ratio of the roughness elements controls the variation of the scaling exponent. The authors also found that with increasing width-to-height ratio the clustering and the lifetime of plumes grow at least in high Pr number fluids, like e.~g. water.

Among this experimental work, there are also a few numerical studies in which thermal convection problems at rough surfaces are addressed. The major challenge of any kind of Direct Numerical Simulation (DNS) is to resolve the boundary layer flow field and, in particular, the fine structure of the flow around the obstacles along with the turbulence in the strongly mixed bulk region. In order to fulfill this requirement, a huge computational effort is needed. Many numerical works are limited to two dimensions. Three of the very recent 2d DNS studies on convection with rough surfaces have been published by Toppaladoddi {\it et al.} \cite{Toppaladoddi2015,Toppaladoddi2017} and Zhu {\it et al.} \cite{Zhu2017}. However, it seems to be not trivial to transfer the results of these simulations to a real 3d geometry, since all kinds of roughness elements generate a strongly three-dimensional flow field near the surface which significantly changes the convective heat flux between the surface and the fluid. One of the rare examples of a 3d DNS has been reported by Stringano {\it et al.} \cite{Stringano2006}. He run a set of DNS for thermal convection at a grooved structure, at whereby the grooves are arranged at the plate in concentric rings. The simulations were performed in a cylindrical RB cell of aspect ratio $\Gam=1/2$ at $Pr=0.7$ and cover a domain in Ra number between $2\times 10^6<Ra<2\times 10^{11}$. The results show an increase of the scaling exponent $\gamma$ if the thickness of the thermal boundary layer becomes smaller than the groove height. Another, very recent numerical work on thermal convection at a non-smooth solid-liquid interface has been published by Wagner \& Shishkina \cite{Wagner2015}. They investigated a cubic RB set-up with four parallelepiped shaped, large obstacles equidistantly attached at the bottom and top plates. The height of the obstacles significantly exceeds the thickness of the thermal boundary layer, which is in strong contrast to the work discussed above. Unlike in all other work, the authors of this study found the increase of the heat flux at their particular surface structure to be lower than the increase of the area caused by the obstacles. However, the total heat flux is still higher compared with a smooth case, and it depends on the height and the distance of the obstacles. The authors evaluated their hypothesis by means of the numerical data up to a maximum Ra number $Ra=10^8$, but they could not validate their data experimentally.

Without raising any claim of completeness of the references above, it can be stated that a great variety of experimental, numerical and theoretical work on thermal convection at rough surfaces has been done. The results of all this work almost coincide in the fact that the enhancement of the heat transfer due to the surface roughness is larger than the pure increase of the surface area (except the work of Wagner \& Shishkina \cite{Wagner2015}). It also seems to be a consensus that the roughness elements induce thermal plumes or even turbulence that increase the mixing in the near-wall flow field. However, there is an obvious contradiction wether or not the scaling exponent in the relation $Nu\sim Ra^{\gamma}$ changes to $\gamma=1/2$, the asymptotic bound predicted for ultra high Ra numbers. Furthermore, there is not yet a universal model meeting all the specific cases of roughness that predicts the convective heat transfer coefficient with an adequate accuracy. An undoubtedly very valuable contribution to derive such a model would be a better knowledge of the local temperature field in the vicinity of the rough surface. This is the main approach of our work. In this spirit, we provide a set of precise and highly resolved temperature data in the parameter domain, where the roughness elements significantly modify the near-wall flow field and we will compare this data with its counterpart from a smooth surface.

\section{Experimental Set-up}
\subsection{Rayleigh-B\'enard cell}

In order to obtain such data in a laboratory experiment, this experiment has to meet two requirements: i) the Rayleigh number has to be sufficiently high to trigger roughness effects in the near wall flow field, ii) the temperature sensor must be much smaller than the characteristic length scales in the boundary layer flow field. These are typically the thickness of the boundary layer or the size of the roughness elements. In the large-scale Rayleigh-B\'enard experiment ``Barrel of Ilmenau'' both requirements are fulfilled very well. A maximum Rayleigh number of $Ra_{max}=10^{12}$ can be set. This is more than one order of magnitude higher than the transitional Ra number at which roughness effects have been observed in the Lyon experiment \cite{Tisserand2011}. At this Ra number, the thickness of the boundary layer amounts to a few millimeters, but the temperature sensor that we used for our measurements has a diameter of only 130~µm, and a length of 330~µm in length. Hence, it is about 50 times smaller than the thickness of the boundary layer and permits a spatial resolution that cannot be reached in any other existing RB cell. The ``Barrel of Ilmenau'' consists of a well insulated container of cylindrical shape with an inner diameter of $D=7.15$~m. It is filled with fresh air. The Prandtl number $Pr=0.7$ remains virtually constant over the entire range of temperatures set during the measurements. A heating plate at the lower side release the heat to the air layer and a cooling plate at the upper side remove it. Both plates are carefully levelled perpendicular to the vector of gravity with an uncertainty of less than 0.15~degrees. The thickness of the air layer $H$ can be varied continuously between $6.30~{\rm m}>H>0.15~{\rm m}$ by moving the cooling plate up and down. The temperature of both plates can be set to values between $20^\circ {\rm C}<T_h<80^\circ {\rm C}$ (heating plate) and $10^\circ {\rm C}<T_c<30^\circ {\rm C}$ (cooling plate). Due to the specific design of both plates (for details see \cite{duPuits2013}), the temperature at their surfaces is very uniform and the deviation does not exceed 1.5~\% of the total temperature drop $\Delta T=T_h-T_c$ across the air layer. The variation of the surface temperature over the time is even smaller and remains below $\pm 0.02$~K. The sidewall of the model room is shielded by an active compensation heating system to inhibit any heat exchange with the environment. Electrical heating elements are arranged between an inner and an outer insulation of 16 and 12 cm thickness, respectively. The temperature of the elements is controlled to be equal to the temperature at the inner surface of the wall. In order to test the efficiency of the system, we set the same temperature of $30.0^\circ \rm{C}$ at the heating and the cooling plates. In case that a heat flux throughout the sidewall exists, the temperature of the interior of the cell would deviate, for outgoing heat towards a lower and for ingoing heat towards a higher temperature. We have measured $29.9^\circ \rm{C}$ indicating that the heat exchange with the environment is very small and can be neglected.

\begin{figure}
  \centerline{\includegraphics[width=10cm]{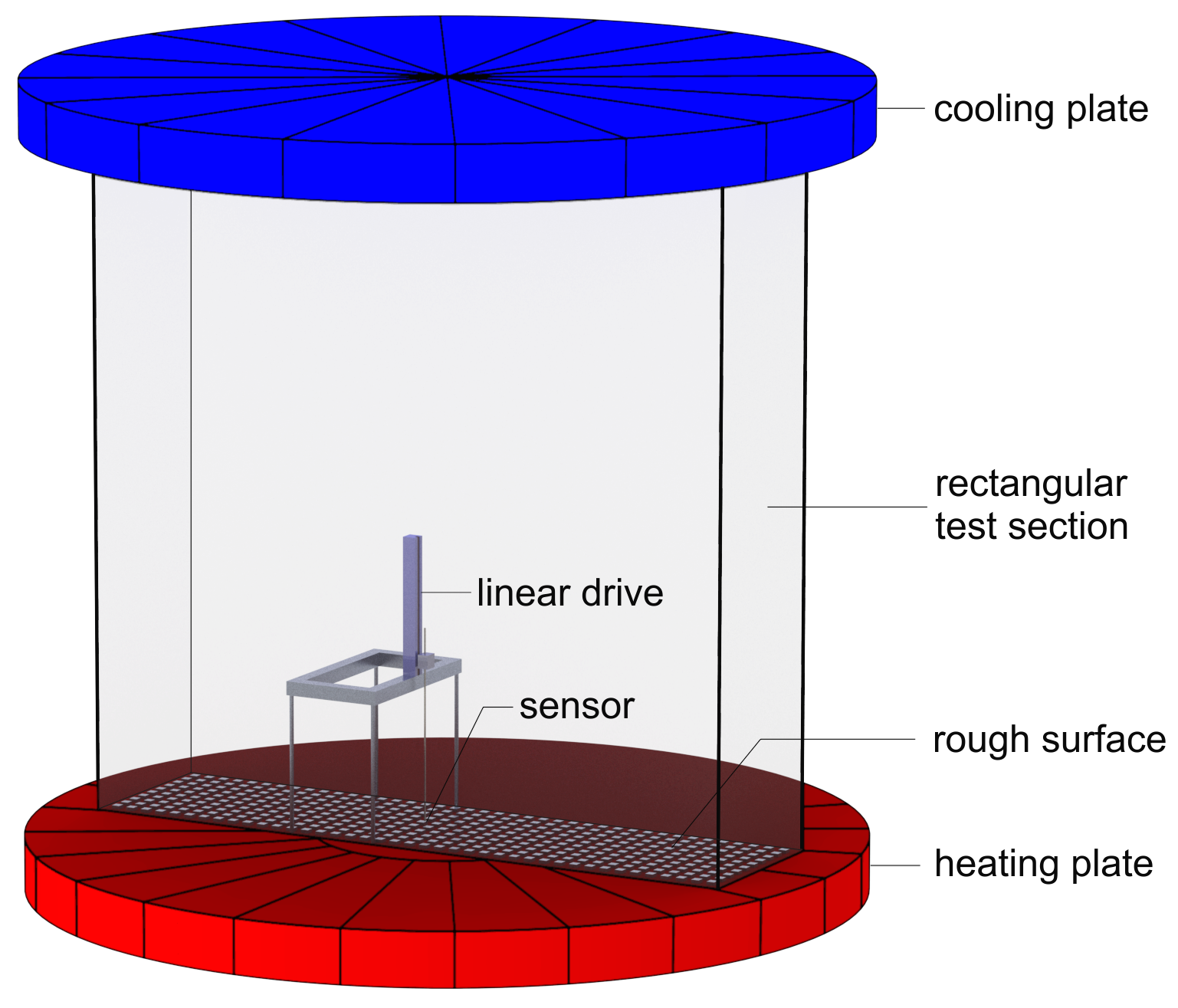}}
  \caption{Rectangular test section to measure the near-wall temperature field in turbulent Rayleigh-B\'enard convection with one rough surface at the heating plate. The test section is build in into the large-scale Rayleigh-B\'enard experiment ``Barrel of Ilmenau''.}
  \label{BOI_Setup}
\end{figure}

Our study of convection reported here was undertaken in a smaller volume of rectangular base area that has been separated from the large test section (see Fig.~\ref{BOI_Setup}). The base area of the inset (defined as x,y-plane) is 2.50~m by 0.625~m, its height (z-coordinate) is 2.50~m. As shown in Fig.~\ref{BOI_Setup}, the original heating and cooling plates of the large cylindrical test cell keep the inset sealed and maintain the stable and well-defined boundary conditions at the bottom and the top of the air layer. We covered the bottom plate with artificial roughness elements, so that our present inset gets actually the same as reported in a recent publication where we investigated the near wall velocity field \cite{Liot2016}. The particular idea behind the slender, cuboid geometry with aspect ratios $\Gam_x=1$ and $\Gam_y=0.25$ is to create a quasi two-dimensional space, in which the orientation of the global flow pattern is fixed with respect to the alignment of the roughness elements. One more important benefit of this configuration with the smaller cuboid RB cell surrounded by the larger cylindrical one is the almost perfect adiabatic boundary condition at the sidewall of the inner test section. This is due to the fact that the vertical temperature profiles $T(z)$ in the large container and in the inner test section (see Fig.~\ref{BOI_Setup} are virtually the same. A heat flux throughout the sidewall of the inner cell is, thus, impossible.

We created the roughness structure at the heating plate by sticking more than 400~little Aluminum blocks at its surface. In order to maintain a uniform temperature of the surface, we took particular care of a good thermal contact between the Aluminum blocks and the surface of the plate. The obstacles with a size of $d=30\times 30~\rm mm^2$ (base area) and $h=12$~mm (height) were arranged in a periodical distance of $l=2d$  as well in x as in y-direction (see Fig.~\ref{Microthermistor Setup}). We specified the height of the obstacles with respect to the typical thickness of the boundary layer in the way that it is smaller at the lowest and larger at the highest Ra number adjustable in our RB cell. The latter quantity has been estimated from the global relation $\delta_{th}=H/2Nu$. The size and the arrangement of the obstacles in the present work is scaled to fit the geometry and the fluid-mechanical similarity with the Lyon experiment \cite{Tisserand2011,Liot2016}. They use a cylindrical cell with a diameter of $D=0.50~\rm m$ and a height of $H=1.00~\rm m$. The working fluid is water with a Prandtl number $Pr=2.5...6.2$. Due to the smaller size, the Lyon experiment is  more flexible with respect to a variation of the boundary conditions (no roughness, various shapes of roughness, single-sided or double-sided roughness). Moreover, it is much better suited to measure the global heat flux precisely and many interesting results has been obtained there. However, it is very difficult to measure the local temperature field in the vicinity of the rough surface with high spatial and temporal resolution. Particularly, this is related to the fact that the boundary layer is typically about 2~mm thick in the Lyon RB cell and even the smallest, water-resistent temperature sensor having a size of about 400~µm, is too large to fully eliminate effects of spatial integration of the temperature field.

\subsection{Temperature measurement technique}

\begin{figure}
  \centerline{\includegraphics[width=12cm]{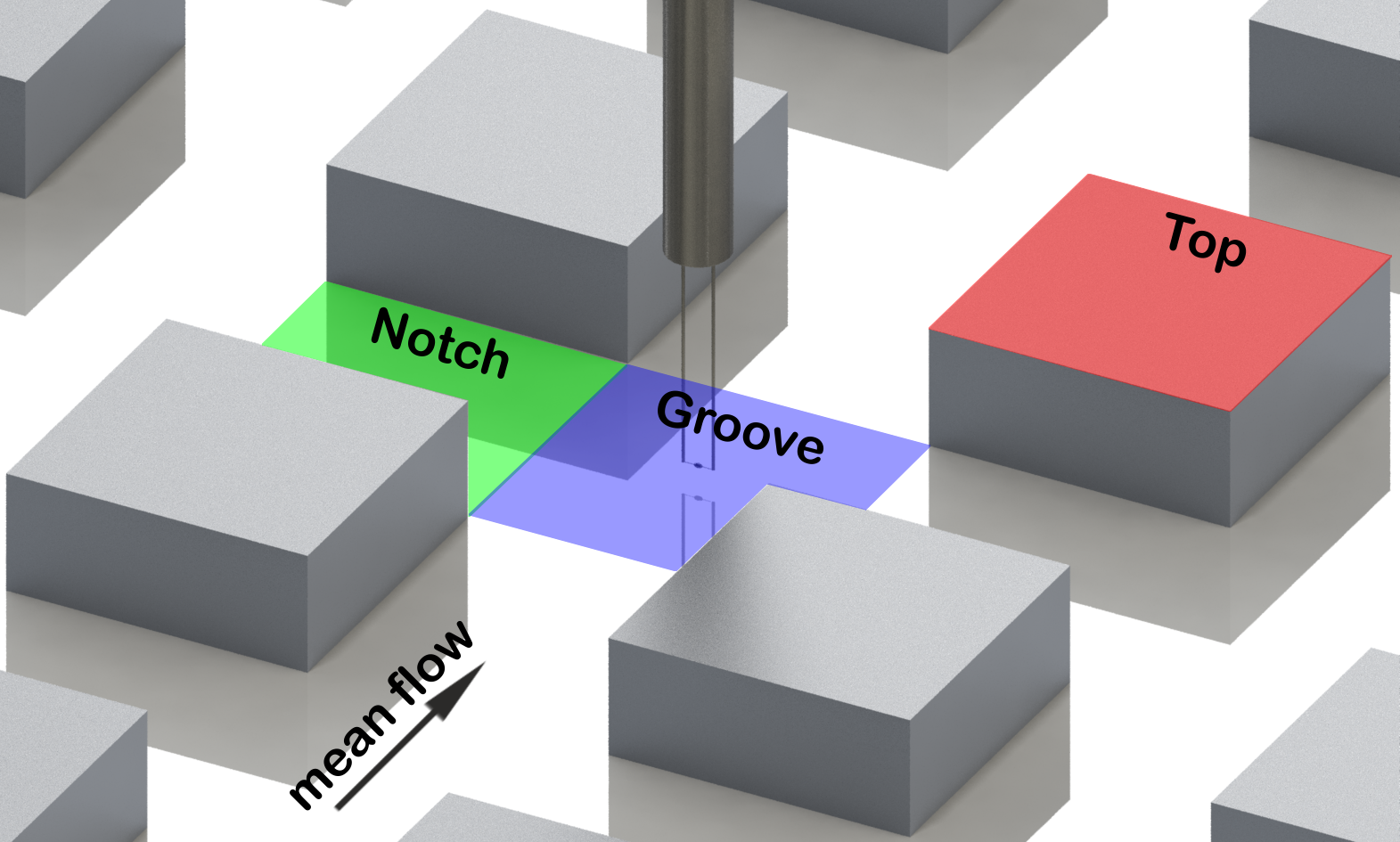}}
  \caption{Arrangement of the artificial roughness elements at the surface of the heating plate with the temperature sensor placed above the Groove area (online: blue). The micro-thermistor is the little dark pearl mounted between the lower tips of the two needles.}
  \label{Microthermistor Setup}
\end{figure}

The measurements reported here will close this gap. As shown in Fig.~\ref{Microthermistor Setup} (true-to-scale), the micro-thermistor that we used to measure the local temperature is much smaller than both, the height of the obstacles and the thickness of the thermal boundary layer, respectively. The temperature sensitive element is the little dark pearl of ellipsoidal shape mounted between the lower tips of the two needles. The typical Reynolds number of the local flow field around the sensor is of the order of $Re_l\approx 1$. This is an important quantity to estimate the effect of the sensor to the near-wall flow field. The thermistor is connected via two 18~µm wires to the conducting support left and right of the sensor. In order to minimize a potential measurement error caused by the strong wall-normal temperature gradients within the thermal boundary layer, the sensor and the connecting wires are aligned parallel to the iso-surfaces of the mean temperature running parallel to the plate surface \cite{Kaiser2012}. The sensor support, a 4~mm rod of brass with the two needles at one side, is mounted at a precise linear motion system with a position accuracy of 0.002~mm. The position at which the sensor touches the surface of the plate, is defined as one half of its diameter or in absolute numbers $z=0.065$~mm. It can be found using a microscope camera.

Another specific requirement on temperature measurements in turbulent flows is the capability of the sensor to track even the fastest temperature fluctuations in the flow. This does not only depend on the properties of the sensor, but also on the characteristics of the flow around it. More precisely, this is mainly related to the size and the advection velocity of the smallest vortices in the turbulent flow. A good estimation for the required resolution in space and time are the Kolmogorov microscales $\eta=(\nu^3/\varepsilon)^{1/4}$ and $\tau_{\eta}=(\nu/\varepsilon)^{1/2}$ with $\nu$ and $\varepsilon$ being the kinematic viscosity and the rate of dissipation of turbulent kinetic energy, respectively. For the lower and the upper end of the Ra number domain of our rectangular test section, $Ra_1=4.6\times 10^9$ and $Ra_2=4.7\times 10^{10}$ - the Kolmogorov length scales amount to $\eta_1=6.4$~mm and $\eta_2=3.0$~mm while the Kolmogorov time scales are $\tau_{\eta,1}=2.7$~s and $\tau_{\eta,2}=0.6$~s, respectively. In order to resolve these scales, the sensor has to be smaller and faster than these numbers. With respect to the size of our thermistor, this demand is fulfilled very well. We also evaluated the response time of the thermistor in a simple laboratory set-up shown in the left sub-figure of Fig.~3. The sensor was placed in a well-defined flow, whose velocity can be set between 0.0~m/s and 1.0~m/s. This is the typical domain of flow velocities close to the rough plate in our convection experiment. Using a Laser beam, we heated the sensor up. Having achieved a steady state, we switched the Laser off, and we measured the decay time. It is quite common to quantify this curve in a single value, at which the sensor has achieved 70~\% of the total jump between the high and the low temperature - the response time $\tau_{70}$.  We follow this and plot the response time of our micro-thermistor in the right sub-figure of Fig.~3. For all flow velocities, even for the case that the surrounding air is in rest, the response time is shorter than the Kolmogorov time scale, and thus, the sensor is capable to resolve even the fastest temperature fluctuations in the boundary layer flow field. In addition to the measurement of the response time, we calibrated the sensor against a primary standard thermometer of PT~100 type with an uncertainty of 0.02~K.

\begin{figure}
  \vspace{0.5cm}
  \makebox[7cm]{\includegraphics[width=5 cm]{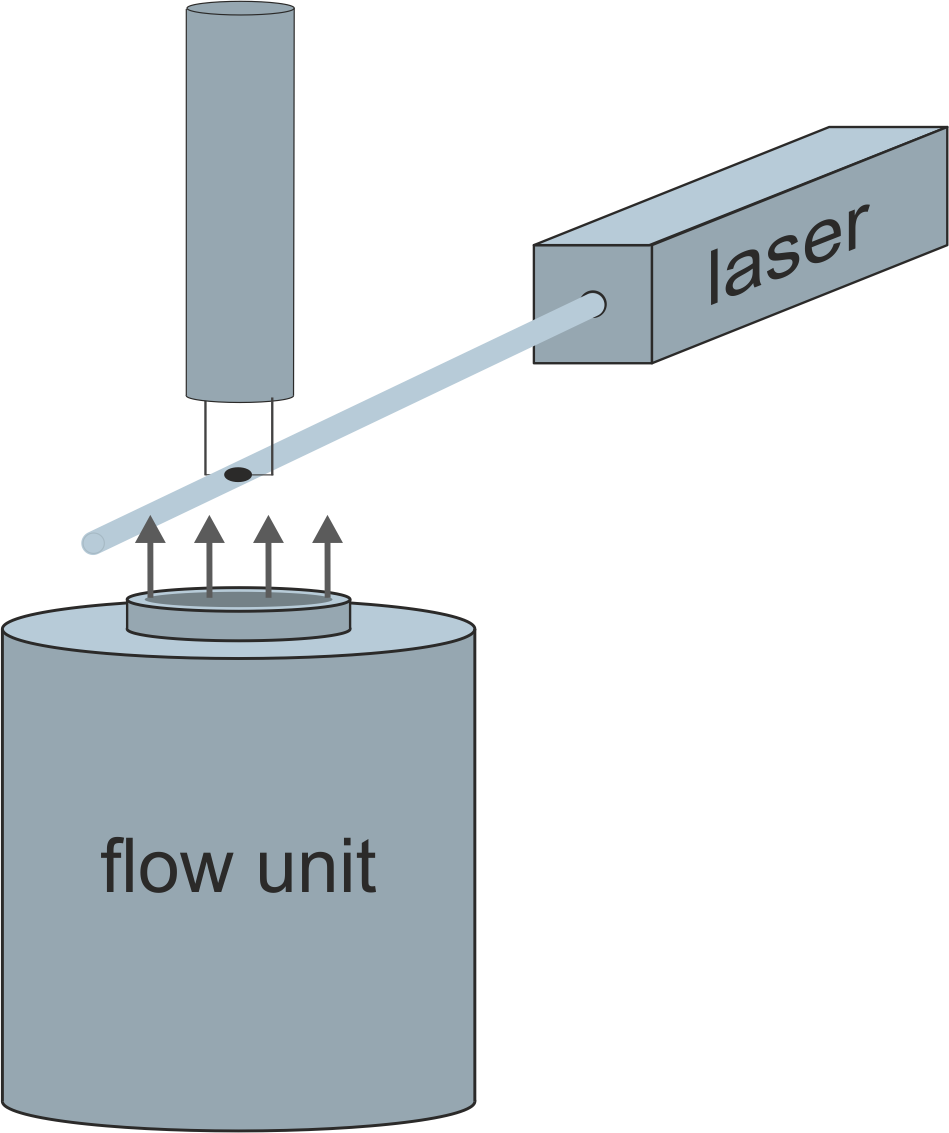}}
  \makebox[7cm]{\includegraphics[width=7 cm]{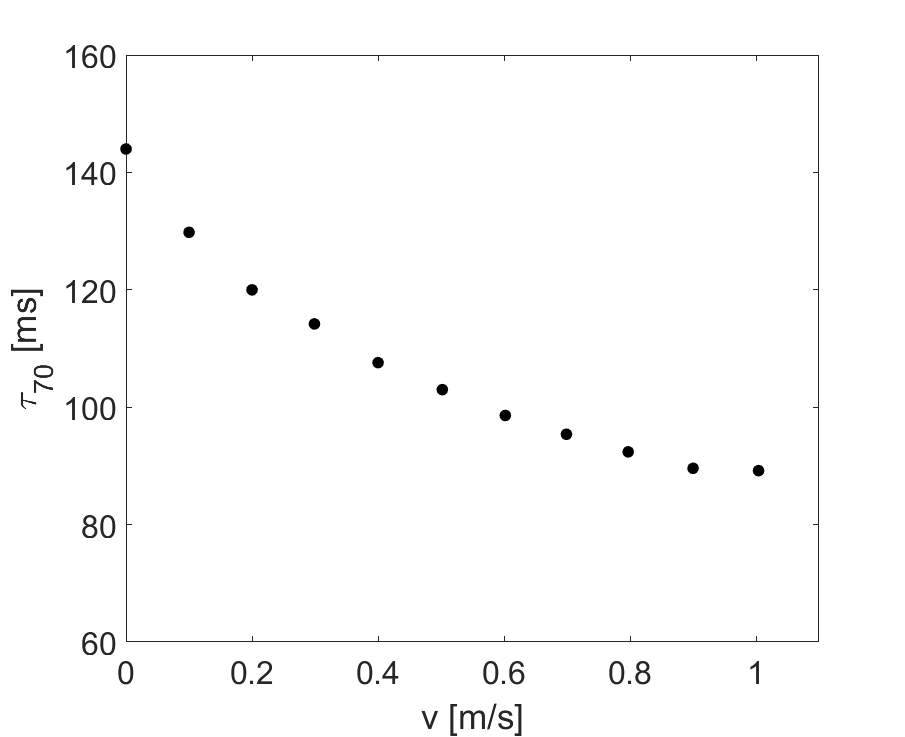}}
\label{response_time}
  \caption{Laboratory set-up to measure the response time of the temperature sensor under well-defined flow conditions (left sub-figure). Measured response time $\tau_{70}$ with respect to the flow velocity (right sub-figure).}
\end{figure}

We also take care of a potential self-heating of the thermistor due to the injected current. We operate the sensor in an active bridge configuration applying a very low sensor current of $I=5\times 10^{-6}$~A. At this current, the measurement error due to the effect of self-heating is estimated to be less than 0.01~K and, hence, below the calibration uncertainty. The output signal of the bridge was acquired by a computer based measurement system with a sampling interval of 5~ms and a dynamic range of 18 bits giving a resolution of the temperature of 5 1/2 digits.

\subsection{Measurement procedure}

Our temperature measurements were undertaken in the same rectangular test section, as recently used to study the velocity field at a rough surface. We chose the same Rayleigh numbers $Ra_1=4.6 \times 10^9$ and $Ra_2=4.7 \times 10^{10}$ corresponding to temperature drops of $\Delta T=3$~K and $\Delta T=40$~K, respectively.  While $Ra_1$ is below transitional effects may start to appear, $Ra_2$ is beyond the transition observed in the global heat flux in the Lyon experiment \cite{Tisserand2011} as well as justified in the near-wall velocity field in the Ilmenau cell \cite{Liot2016}. We have undertaken our present temperature measurements at the centre of the heated bottom plate. With respect to the installed roughness pattern and the orientation of the mean flow, we define three distinct surface regions in our set-up (cf. Fig.~\ref{Microthermistor Setup}) that we refer to as: i) ``Grooves''~--~these are the valleys in which the air flow is virtually not disturbed by the obstacles, ii) ``Notches''~--~these are the surface areas in between the obstacles where the obstacles block the flow, and iii) ``Top''--these are the upper surfaces of the obstacles. We measured the temperature profiles $T(z)$ at each of these surface regions, considering the strong variations that have been reported with respect to the measured velocity field \cite{Liot2016}. Each of the profiles was measured point-by-point, moving the sensor in a vertical line away from the surface of the plate. It is assembled from 35 different z-positions  starting at $z=0.065$~mm ($z/H=2.6 \times 10^{-5}$) and ending up at $z=147.66$~mm ($z/H=0.059$). Time series of 540,000 samples have been acquired over a period of 2700~s at every single position providing sufficient data for an adequate statistical convergence. In the subsequent sections, we will discuss the data and we will derive a few more general conclusions that describe the local heat transport in turbulent convection with rough surfaces.

\section{Results}
\subsection{Statistical analysis of the temperature field}

One of the unquestionable results of all experimental work on convection at rough surfaces is the fact that the increase of the heat flux is larger than the increase of the surface due to the roughness elements (see e.~g. Du \& Tong \cite{Du2000} or Ciliberto \& Laroche \cite{Ciliberto1999}). Our highly resolved temperature measurements that start very close to the surface of the heating plate enable us, to investigate the local temperature field and to quantify the local heat transport at distinct surface areas. To this aim, we measured temperature profiles at all three surface regions ``Groove'', ``Notch'' and ``Top'' (cf. Fig.~\ref{Microthermistor Setup}), and we compare the results with earlier measurements at a smooth surface. The latter measurements have been undertaken in two different set-ups: i) in the full-size ``Barrel of Ilmenau'' with a diameter of $D=7.15~{\rm m}$ and a height of $H=6.30~{\rm m}$ for the higher Ra number $Ra=5.2\times 10^{10}$, and ii) in a smaller inset of cylindrical shape with a diameter of $D=2.50~{\rm m}$ and a height of $H=2.50~{\rm m}$ for the lower Ra number $Ra=3.4\times 10^{9}$. The working fluid is also air with a Pr number of $Pr=0.7$ in these two experiments. The sensor and the measurement technique used for the measurements at the smooth surface were the same as applied at the rough one. We wish to refer here to two references \cite{Li2012,duPuits2013}, where the reader will find more details on the facility as well as the original data from both, the lower and the higher Ra number measurements.

\begin{figure}
  \centerline{\includegraphics[width=15cm]{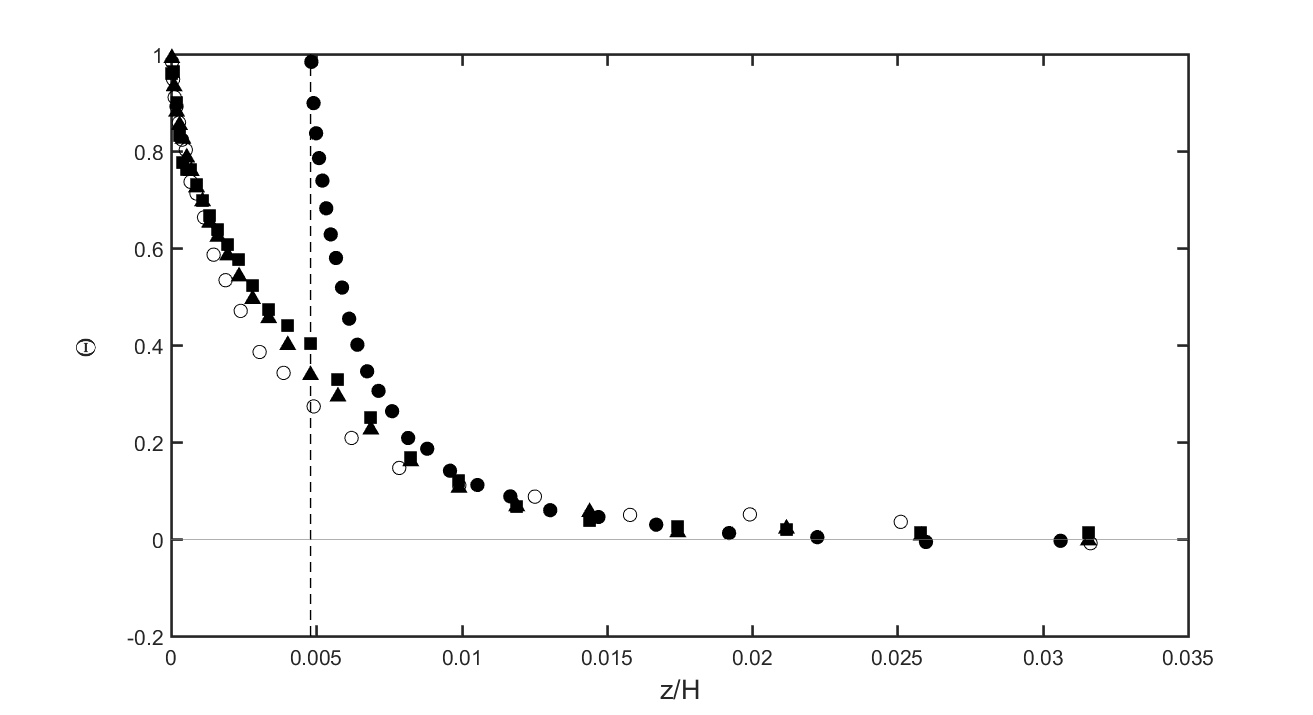}}
  \caption{Low Rayleigh number: Profiles of the mean temperature $\Theta(z/h)$ measured at the rough surface at  $Ra_1=4.6 \times 10^9$, Top: black circles, Notch: black squares, Groove: black triangles, and at the smooth surface ($Ra=3.4 \times 10^9$): open circles. The dashed vertical line at $z/H=4.8 \times 10^{-3}$ indicates the height of the obstacles.}
  \label{Tmean_lowRa}
\end{figure}

\begin{figure}
  \centerline{\includegraphics[width=15cm]{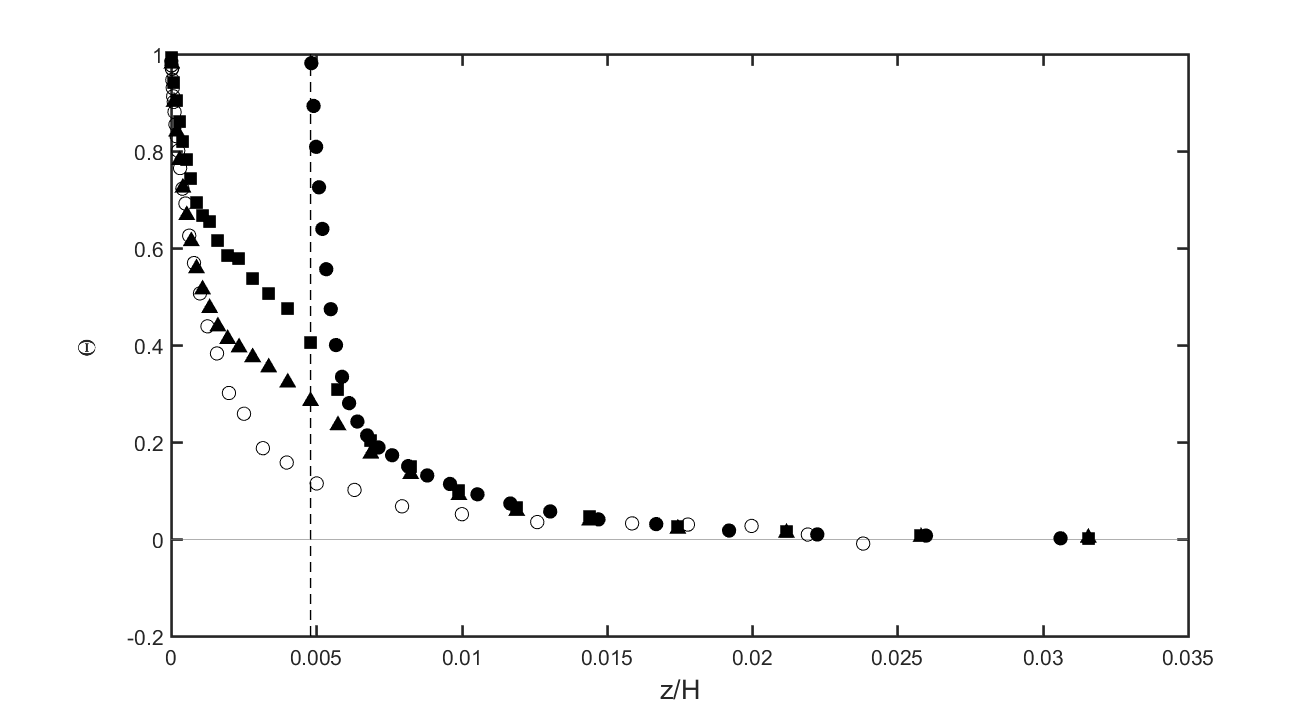}}
  \caption{High Rayleigh number: Profiles of the mean temperature $\Theta(z/h)$ measured at the rough surface at $Ra_2=4.7 \times 10^{10}$, Top: black circles, Notch: black squares, Groove: black triangles, and at the smooth surface ($Ra=5.2 \times 10^{10}$): open circles. The dashed vertical line at $z/H=4.8 \times 10^{-3}$ indicates the height of the obstacles.}
  \label{Tmean_highRa}
\end{figure}

We start our discussion with the profiles of the normalized mean temperature $\langle \Theta\rangle$:
\begin{equation}
\langle \Theta\rangle=(\langle T(t)\rangle -T_B)/(T_H-T_B)
\label{mean_temperature}
\end{equation}
versus the normalized distance $(z/H)$ from the plate surface that are plotted in Fig.~\ref{Tmean_lowRa} for $Ra_1=4.6 \times 10^9$ and in Fig.~\ref{Tmean_highRa} for $Ra_2=4.7 \times 10^{10}$. Here, the temperature $T(t)$ represents the temperature time series measured using the micro-thermistor at the various distances $z$ from the surface of the plate. $T_B$ is the so-called bulk temperature measured using a separate Resistance Temperature Device (RTD) of PT~100 type in the center of the cell which is well mixed and of virtually uniform temperature. $T_H$ is the temperature of the heating plate also measured using a RTD within the plate. In order to plot all profiles in the correct geometric relation, we indicate the height of the obstacles as a dashed vertical line at $z/H=4.8 \times 10^{-3}$, and we shift the profile of the mean temperature at the top of the obstacles by this height. At both Ra numbers, the temperature gradient at the top of the obstacles, which is a direct measure of the local heat flux density according to Fourier's law
\begin{equation}
\dot{q}=-\lambda ~\partial \langle T \rangle / \partial z|_{z=0},
\label{fouriers_law}
\end{equation}
significantly exceeds as well the gradients in the Groove and the Notch as that gradient measured at the smooth surface. This is a first conclusion of generality, the heat flux enhancement observed at rough surfaces is mainly accounted for by an increase of the local heat transfer coefficient at the top of the obstacles. And, this is true below and beyond the critical Ra number, at which a transition in the scaling exponent $\gamma$ was observed. For completeness, we have listed the numbers in Table~\ref{Gradients}.

The transition of the scaling exponent $\gamma$ beyond a critical Ra number $Ra_c$ \cite{Ciliberto1999,Roche2001, Qiu2005} represents as well an interesting and not well understood feature of thermal convection at rough surfaces. This is quite similar to RB convection with smooth plates, however, the transition starts at a much lower $Ra_c$. Schumann provided a first approximation to predict its onset \cite{Schumann1988}. Applying Eq.~\ref{limitsmallrough} to our measurements, we obtain a critical Ra number $Ra_c=560~(2500~{\rm mm}/12~{\rm mm})^{8/3}=8.5 \times 10^8$ as the bound for this transition. In an alternative manner, Du \& Tong  predicted the onset of roughness effects at all, if the thickness of the boundary layer $\delta_T$ falls below the height of the obstacles $h$ \cite{Du2000}. Applying this prediction to our experiment, it means that $\delta_T$ must fall below $h=12$~mm. This appears at a critical Nu number $Nu_c=H/2h=2500~{\rm mm}/24~{\rm mm}=104$. Since comprehensive data of the scaling $Nu\sim Ra^{\gamma}$ at $Pr=0.7$ is not available for a rectangular box of two different aspect ratios and rough top and bottom plates, we estimate the corresponding Ra number using the Grossmann-Lohse model that is developed for smooth boundaries \cite{Grossmann2001}. We obtain $Ra_c=5.6\times 10^9$, which is only about half a decade higher than Schumann's prediction. From the work by \cite{Liot2016}, we also have a prediction, when roughness effects start to modify the heat transport in our specific test section of rectangular shape. The authors of this work investigated the velocity field in the Notch region and they found the fluid being confined within the Notch below a critical Ra number, while there is a fluid exchange with the bulk flow beyond it. They also analysed the scaling of the the local wall heat flux that they measured directly using heat flux sensors. From both measurements, they found a consistent limit for the onset of roughness effects which is as high as $Ra_c=1.5\times10^{10}$. This is in a fair agreement with the two predictions by Schumann and Du\&Tong, and the little differences with respect to those models might be a consequence of the finite aspect ratio as well as specific geometry of the rectangular test section with aspect ratios $\Gam_x=1$ and $\Gam_y=0.25$.

\begin{figure}
  \centerline{\includegraphics[width=15cm]{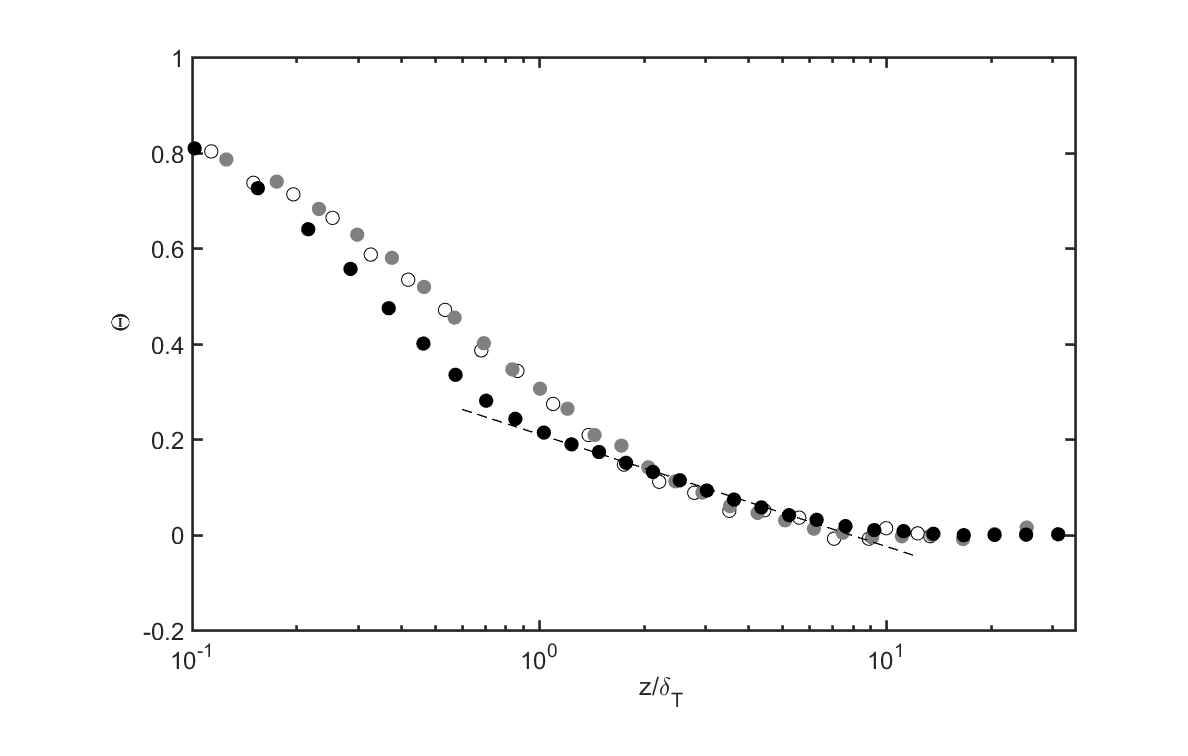}}
  \caption{Profiles of the mean temperature above the top of the obstacles at $Ra_1=4.6 \times 10^9$ (gray circles) and $Ra_2=4.7 \times 10^{10}$ (black circles) compared with the profile at the smooth plate at $Ra=3.4 \times 10^9$ (open circles). The dashed line fits the profile at the higher Ra with a logarithm according to $\Theta=0.102~{\rm log}(z/\delta_T)+0.21$.}
  \label{Mean_top_comparison}
\end{figure}

In the work presented here, we wish to find out whether or not the transition observed in the scaling of the local heat transport as well as in the local flow field in the Notch can be attributed to a variation in the local temperature field and can we learn something more from our highly resolved temperature measurements. To this end we plot the profiles of the mean temperature at the Top of the obstacles in a semi-logarithmic manner (see Figure~\ref{Mean_top_comparison}). The profiles are scaled by the various thicknesses of the boundary layer at $Ra_1$ and $Ra_2$ (see Table~\ref{Gradients}) and thus, they are directly comparable. We used the displacement thickness for this normalization that is defined as \begin{equation}
\delta_{T}=\int_0^{z_{max}} \left\{1-\overline{\Theta}(z)\right\}dz.
\label{eq:displacement_thickness}
\end{equation}
It is quite obvious that the profile at the higher Ra number significantly differs from those two at the lower Ra number and at the smooth plate, respectively. The curve is steeper and in the flow region $1<z/\delta_T<10$, it exhibits a logarithmic behavior. In particular, the latter fact is a typical attribute of a turbulent boundary layer. In fact, the roughness elements shift the laminar-turbulent transition of the boundary layer and, hence, the transition of the scaling exponent towards lower Ra numbers with respect to the prediction in RB convection with smooth plates \cite{Grossmann2001}. However, it should be noted here as well that the variation of the scaling exponent $\gamma$ is expected to be smaller with respect to the transition in RB convection with smooth surfaces since the enhancement of the local heat flux concerns only the Top region and, thus only one forth of the total surface area.

\begin{table}
\begin{center}
\begin{tabular}{lcccc}
&                     ~~  $\partial \langle \Theta \rangle / \partial (z/H)|_{z/H=0}$ &~~ $\partial \langle T \rangle /\partial z ~[\rm{Km^{-1}]}$ &~~ $\delta_T/H~ [\times 10^{-3}]$ & ~~$\delta_T ~\rm{mm}$ \\[3pt]
$Ra_1=4.6 \times 10^9$ \\
Groove & 562.1 & 255 & 4.394 & 10.99  \\
Notch & 514.9 & 238 & 4.691 & 11.72  \\
Top & 784.9 & 340 & 2.326 & 5.82  \\
smooth & 480.0 & 225 & 4.473 & 11.18\\[3pt]
$Ra_2=4.7 \times 10^{10}$ \\
Groove & 848.9 & 4406 & 3.461 & 8.65 \\
Notch & 494.6 & 2790 & 4.466 & 11.17 \\
Top & 1048.2 & 5500 & 1.888 & 4.72 \\
smooth & 824.7 & 170 & 2.445 & 15.40 \\
\end{tabular}
\end{center}
\caption{Gradients of the mean temperature and (displacement) thickness of the thermal boundary layer at the distinct areas at rough and smooth surfaces for two Rayleigh numbers. (Please note that the dimensional temperature gradient as well as the dimensional thickness of the thermal boundary layer at the smooth plate ar $Ra_2=4.7 \times 10^{10}$ are not fully comparable with the other numbers since the total height of this experiment was $H=6300~\rm{mm}$ instead of $H=2500~\rm{mm}$) for all other measurements.}
\label{Gradients}
\end{table}

The interpretation of the measured mean temperature profiles in the Groove and the Notch regions is not that easy and it seems to be hard to predict a quantitative effect of the observed modification in the flow field to the scaling of the global heat transport. For the lower Ra number $Ra_1=4.6 \times 10^9$, all profiles including the temperature gradients at the wall collapse fairly well with each other and with the profile at the smooth plate (except of course that profile measured above the Top). For the higher Ra number $Ra_2=4.7 \times 10^{10}$, the profiles of the mean temperature exhibit clear differences. In particular, in the region $0.001<z/H<0.048$, the latter number corresponds with the height of the roughness elements, those profiles are more flat compared with the reference at the smooth plate. This indicates an enhancement of the convective heat transport $\langle w'T'\rangle$, but not necessarily a transition to a turbulent state of the boundary layer in the Groove and in the Notch regions. In order to explain these variations of the temperature profiles, it might be useful to have again a look back into the recent study of the velocity field undertaken in the same RB cell with one rough surface \cite{Liot2016}. The authors of this paper report a transition of the flow field in the Notch, when the Ra number exceeds about $Ra_c=1.5\times10^{10}$. For more clearness of the reader, we re-plot Figs.~4 and 5 of that paper here. The figures show that the fluid remains confined in the Notch below $Ra_c$, and it forms a little RB cell with a local Ra number of $Ra_{l1}\approx 100$. This number is based on the height of the obstacles and the vertical temperature drop across the Notch. It considerably remains below the stability limit $Ra_s=1707$, at which Lord Rayleigh  and Jeffreys predicted the onset of convection in an laterally infinite fluid layer \cite{Rayleigh1916,Jeffreys1926}. However, we found little temperature fluctuations in the Notch (to be discussed along with the profiles of the standard deviation below), which indicates that heat is not only transported by diffusion. A little fraction of convective transport also takes place at the lower Ra number. Beyond $Ra_c$ the situation in and around the Notch completely changes. The fluid is no longer trapped inside the Notch and starts to leave it. This induces a much stronger convective transport process between the fluid in the Notch and the fluid above the obstacles and significantly enhances the efficiency of the heat transport. This becomes also visible in the gradient of the mean temperature, which is much smaller than below the transition. The process, that which enables the fluid to leave the Notch at the higher Ra number, is mainly triggered by local shear forces due to the mean wind, but it is also supported by local buoyancy forces within the notch. The latter ones can be estimated again computing the local Ra number which amounts to $Ra_{l2}\approx 1100$. This value is still below the stability limit. However, the velocity measurements show that both contributions together are strong enough to trigger the exchange of fluid between the Notch and the mean flow. The temperature profile in the Groove at the higher Ra number is also more flat than that at a smooth surface. For this observation, we do not have an explicit explanation. However, it seems to be very likely that the transition of the flow field in the Notch also affects the flow field in the Groove.
\begin{figure}
  \centerline{\includegraphics[width=13cm]{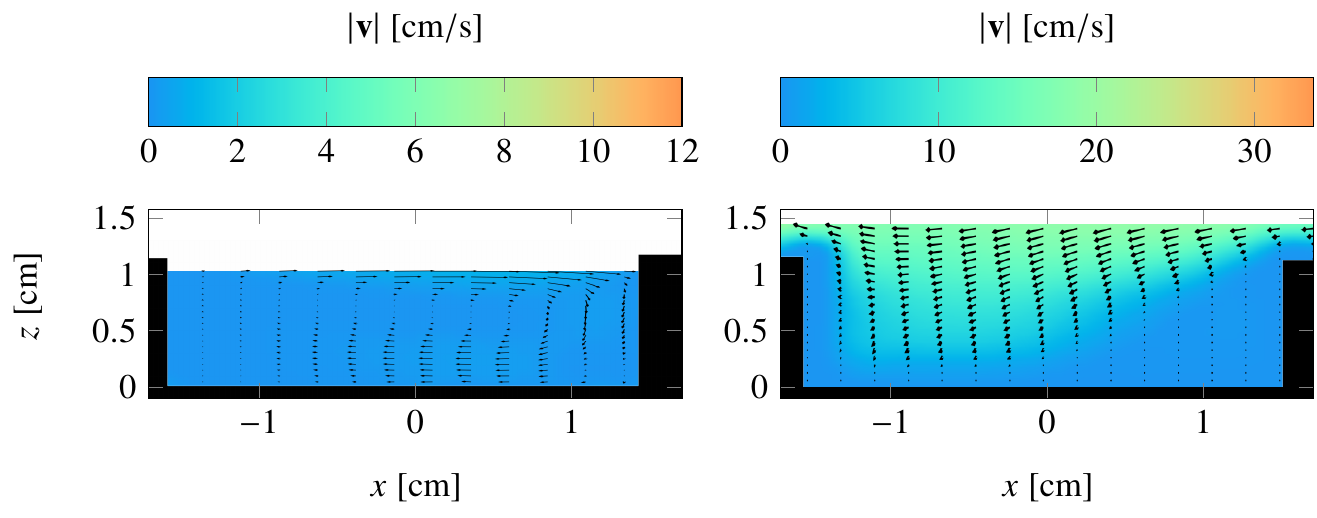}}
  \caption{Mean velocity fields inside the Notch at $Ra=4.7 \times 10^9$ (left subfigure) and $Ra=4.0 \times 10^{10}$ (right subfigure). The scale of the arrows is arbitrary and differs from one plot to another to allow better visualization of the flow \cite{Liot2016}}
    \label{Mean_velocity_field}
\end{figure}
\begin{figure}
  \vspace{0.5cm}
  \centerline{\includegraphics[width=12cm]{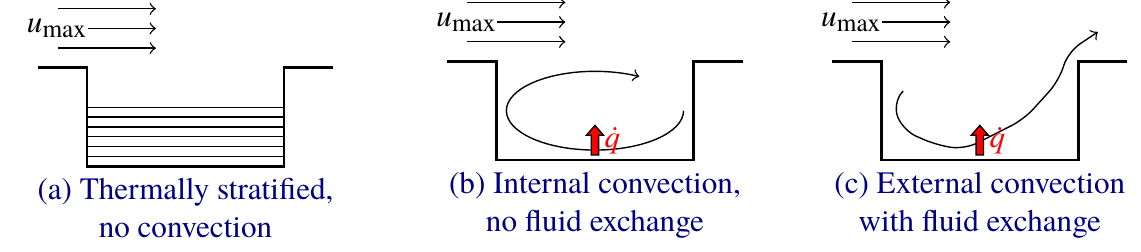}}
  \caption{Sketch of possible flow structure inside a Notch. (a) Thermally stratified, no convection, (b) internal convection, no fluid exchange, (c) external convection with fluid exchange (original Fig.~5 from \cite{Liot2016}) }
  \label{Potential_flow_structures}
\end{figure}

\begin{figure}
  \centerline{\includegraphics[width=15cm]{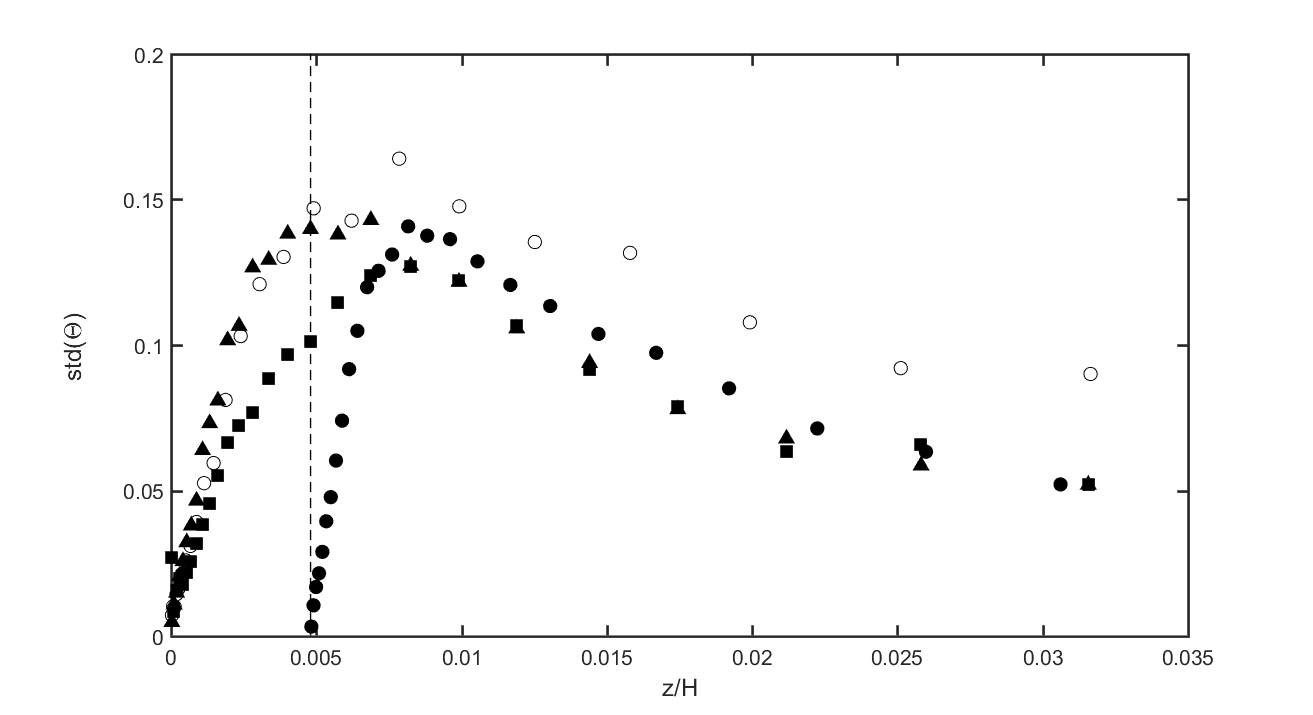}}
  \caption{Low Rayleigh number: Profiles of the normalized standard deviation $\rm{std}(\Theta(z/H))$ at various locations of the rough surface ($Ra_1=4.6 \times 10^9$), Top: black circles, Notch: black squares, Groove: black triangles, and at the smooth surface ($Ra=3.4 \times 10^9$): open circles. The dashed vertical line at $z/H=4.8 \times 10^{-3}$ indicates the height of the obstacles.}
  \label{Tstd_lowRa}
\end{figure}

\begin{figure}
  \centerline{\includegraphics[width=15cm]{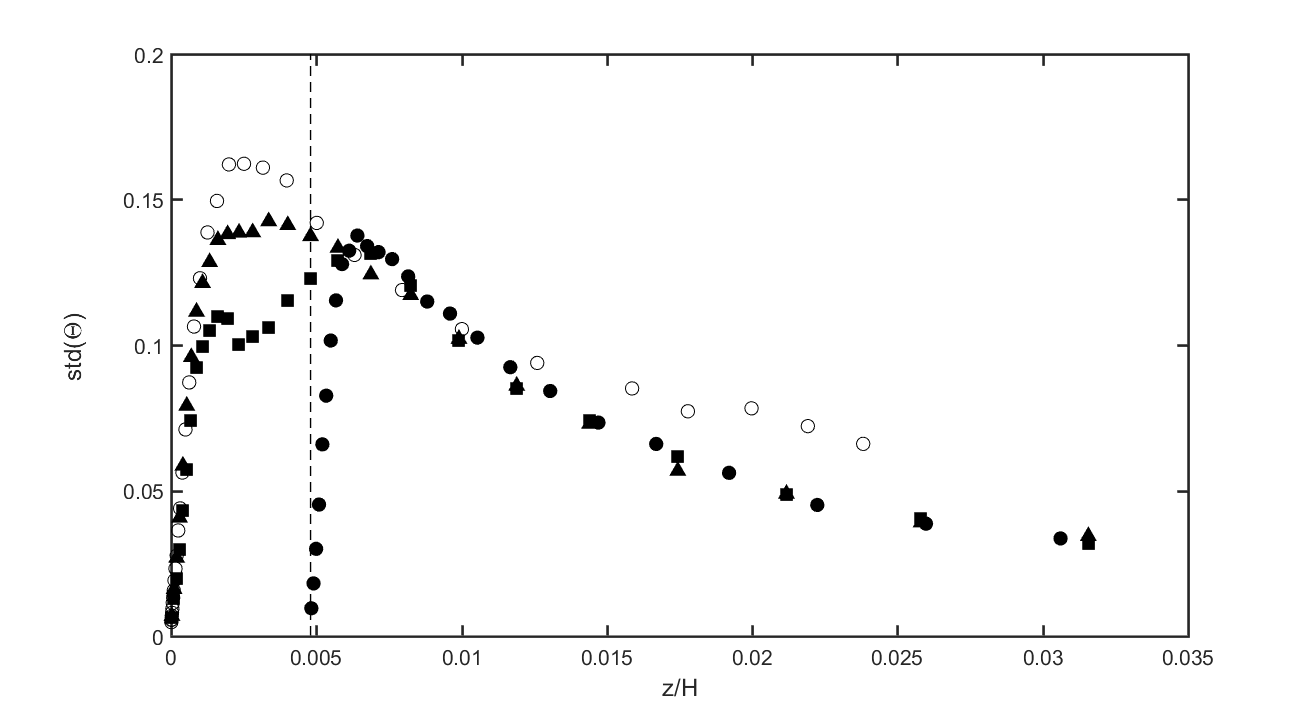}}
  \caption{High Rayleigh number: Profiles of the normalized standard deviation $\rm{std}(\Theta(z/H))$ at various locations of the rough surface ($Ra_2=4.7 \times 10^{10}$), Top: black circles, Notch: black squares, Groove: black triangles, and at the smooth surface ($Ra=5.2 \times 10^{10}$): open circles. The dashed vertical line at $z/H=4.8 \times 10^{-3}$ indicates the height of the obstacles.}
  \label{Tstd_highRa}
\end{figure}

In a next step, we are going to discuss the fluctuations, respectively the standard deviation of the near-wall temperature field. The fluctuations are a specific signature of thermal plumes are believed to contribute significantly to the heat transport process. We compute the standard deviation of the temperature time series, and we normalize the result by the temperature drop between the surface temperature of the heating plate and the temperature in the bulk:
\begin{equation}
std (\Theta) = \frac{1}{T_H-T_B}\sqrt{\frac{1}{N-1}\sum_{i=1}^N (T_i-\langle T\rangle)^2},
\end{equation}
We plot the profiles of the normalized standard deviation for both Ra numbers in the diagrams of Fig.~\ref{Tstd_lowRa} and Fig.~\ref{Tstd_highRa}, respectively. Again, the profiles at the top of the obstacles are shifted by $\Delta(z/H)=4.8 \times 10^{-3}$. Since all profiles are scaled by the total temperature drop between the heating plate and the bulk, they can be directly compared. In order to analyse some potential variation of the flow with respect to the smooth case, we have added profiles of the temperature fluctuations over a smooth surface.

We start discussing the lower Ra number case at $Ra_1=4.6 \times 10^9$ (see Fig.~\ref{Tstd_lowRa}). The profile of the temperature fluctuations in the Groove collapse very well with that measured at the smooth reference. This is not surprising and requires no further discussion, since the flow situation in the Groove is quite comparable with the flow along a smooth plate. In contrast to this, the fluctuations in the Notch are generally smaller, which can be explained by the confinement of the fluid in between the obstacles and the decoupling of this fluid volume from the fully turbulent mean wind (see Figs.~\ref{Mean_velocity_field} and \ref{Potential_flow_structures}). Beyond the height of the obstacles at $(z/H)>4.8 \times 10^{-3}$ the temperature fluctuations at all specific regions of the rough surface are generally smaller than those at the smooth surface. This sounds a bit mysterious, since due to the obstacles with all their corners and edges, we rather expected an opposite behavior. One potential explanation might be the different geometry of the rough and the smooth cells. While the flow is locked in a single direction in the rough (rectangular) cell, it has the freedom to change its orientation in the smooth (cylindrical) one. The flow in the latter case is, thus, more complex and larger velocity fluctuations are very likely. This must be reflected in the temperature field as well since both fields are strictly coupled. Insofar, the observation discussed above, might rather be associated with the shape of the cell than with the existence of roughness at the plate surface.

For the higher Ra number case, meaning beyond the transition in the global $Nu(Ra)$ relation, the profiles of the temperature fluctuations clearly differ from those measured at the lower Ra number case case. The maximum of the fluctuations in the Groove is shifted from a position above the height of the obstacles to a position in between them, indicating that the thickness of the boundary layer really becomes smaller than the height of the obstacles. In this content, it is also interesting to point out that the fluctuations in the Notch exhibit a buckle at $z/H=1.8 \times 10^{-3}$. This kink as well as the steeper increase of the profile very close to the wall are consequences of the onset of the flow exchange within the Notch and the turbulent mean wind above the roughness elements (cf. Fig.~\ref{Mean_velocity_field}). With respect to the convective term $\langle w'T'\rangle$, these are also indicators of a generally more efficient heat transport than in the case of the lower Ra number.


A final interesting aspect that we are able to discuss in this section thanks to our highly resolved temperature measurements, is the question, at which distance from the plate the flow ``forgets'' the specific geometry of the surface. With respect to the profiles of the mean temperature and the temperature fluctuations we identify this as the z-position, at which the ``rough'' profiles collapse with the ``smooth'' case. Considering all the curves in Figs.~\ref{Tmean_lowRa},~\ref{Tmean_highRa},~\ref{Tstd_lowRa}, and \ref{Tstd_highRa}, this point can be roughly identified with a specific distance of twice the height of the obstacles. Insofar, it is undoubtedly verified that the the enhancement of the convective heat flux due to an artificial roughness of the solid surface is, at least in the domain of Ra numbers investigated here, associated with a modification of the near-wall temperature field.

\subsection{Temperature fluctuations and plume dynamics}

\begin{figure}
  \centerline{\includegraphics[width=14cm]{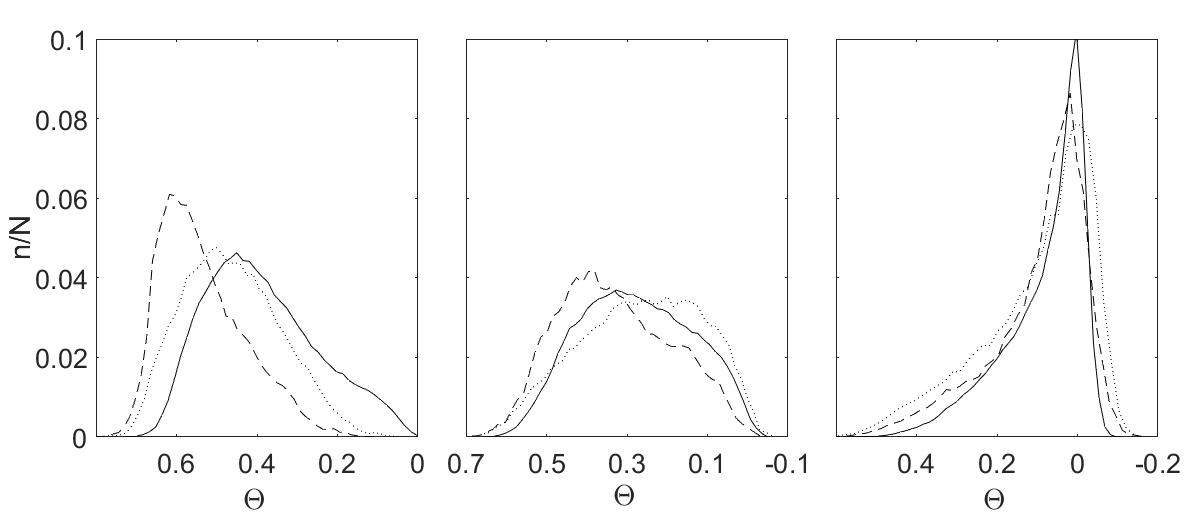}}
  \centerline{\includegraphics[width=14cm]{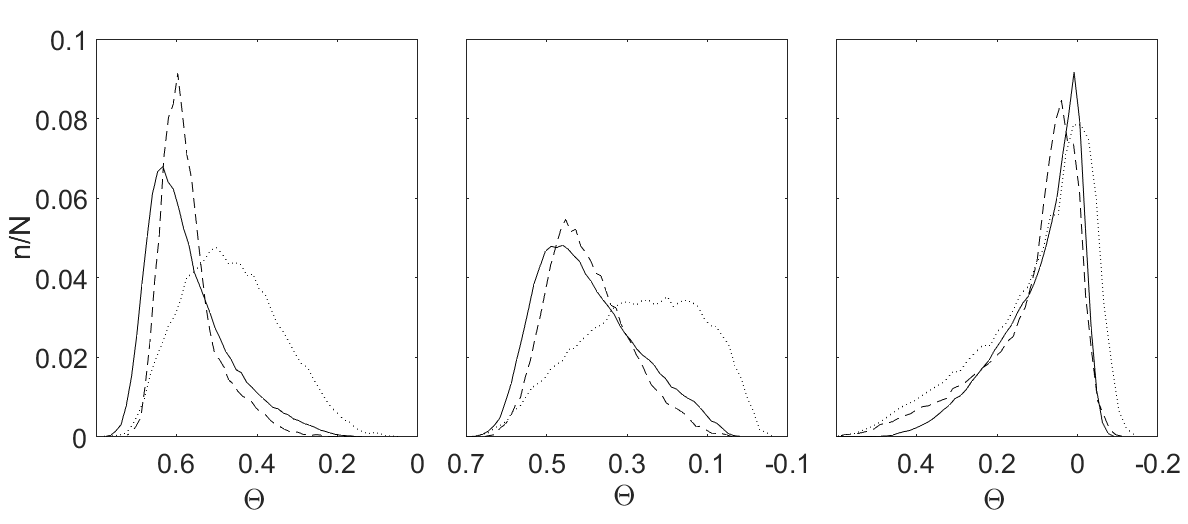}}
  \centerline{\includegraphics[width=14cm]{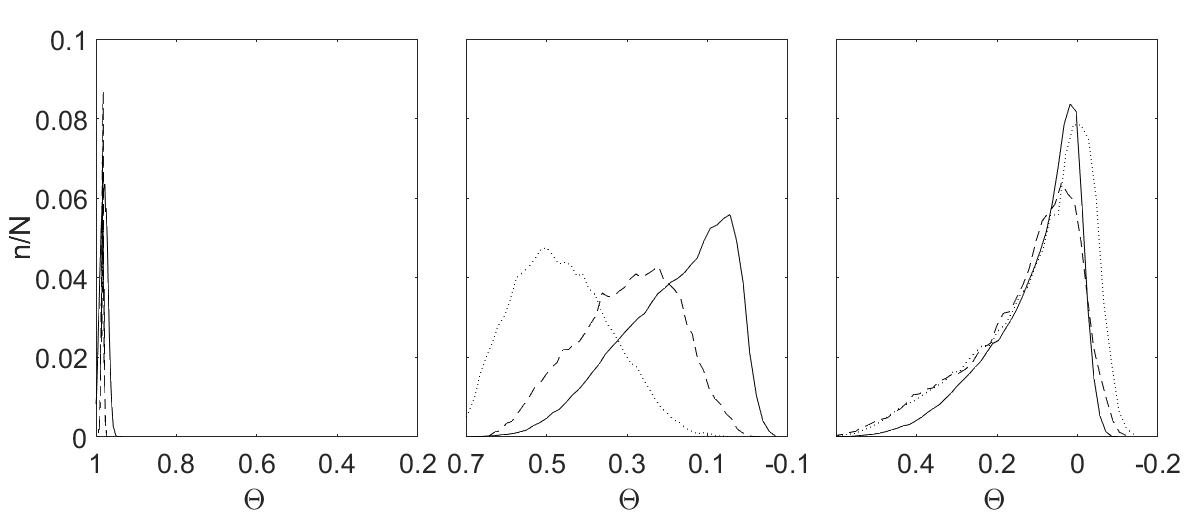}}
  \vspace{0.1cm}
  \caption{Histograms of the samples of the normalized temperature $\Theta(t)$ at various locations of the rough surface. In all plots, the full and the dashed lines represent the rough case at $Ra_2=4.7 \times 10^{10}$ and $Ra_1=4.6 \times 10^9$, respectively. For comparison, we have added the smooth case at $Ra=3.4 \times 10^9$ (dotted line). Upper row: Groove, intermediate row: Notch, lower row: Top. From left to right $z/H=0.00234$, $z/H=0.0048$, and $z/H=0.0099$ ($z/H=0.0048$, $z/H=0.0071$, and $z/H=0.0096$ for the lower row).}
  \label{Histograms}
\end{figure}

In the preceding two sections, we have discussed some statistical quantities of turbulent RB convection with rough surfaces and we have compared the results with earlier measurements at smooth surfaces. In this section, we wish to investigate the particular properties of the temperature fluctuations, that are a signature of the evolution of thermal plumes at the roughness elements. In particular close to the wall, where a strong, wall-normal temperature gradient appears, the fluctuations directly reflect the underlying flow field. In so far, their analysis contributes to understand the specific dynamics of the evolution of thermal plumes and the turbulent heat transport in the near-wall flow region. Based on the discussion above and taking into account some observations from previous work we wish to check a few particular quantities that potentially indicate a transition of the near wall flow field from a fluctuating, but still rather laminar to a turbulent flow state. These aspects are designated in the following list.
\begin{list}{•}{
\setlength{\leftmargin}{0.45cm}
  \setlength{\itemindent}{-0.45cm}}
  \item ~~There is a transition of the flow field in the Notch if the the Rayleigh number is increased from $Ra_1=4.6 \times 10^9$ to $Ra_2=4.7 \times 10^{10}$ (see Fig.~\ref{Mean_velocity_field}).
  \item ~~A logarithmic region in the mean temperature profile above the top of the obstacles emerges at the higher Rayleigh number $Ra_2=4.7 \times 10^{10}$ (see Fig.~\ref{Mean_top_comparison}).
  \item ~~In the Notch, the profile of the temperature fluctuations exhibits a ``buckle'' at $Ra_2=4.7 \times 10^{10}$ (cf. Fig.~\ref{Tstd_highRa}).
  \item ~~There is a hypothesis that the frequency and the dynamics of the plume emission at a rough surface changes with respect to a smooth one (see \cite{Du2000}).
\end{list}
First, we analyse the probability density function of the temperature fluctuations. To this aim, we plot selected histograms of the normalized temperature $\Theta(t)$ in Fig.~\ref{Histograms}. In all plots, the dashed lines represent the lower Ra number case below the transitional effect sets in, the full lines represent the higher Ra number case beyond the transition, and the dotted lines represent the distribution of the samples at a smooth surface at $Ra=3.4 \times 10^9$. For reasons of clarity, we only show the lower Ra number case for the smooth surface, since it equals the higher Ra number case. All histograms are partitioned into 50 bins and they are normalized by the total number of samples $N$ within a single time series. For a more clear representation, we fitted the discrete data points of the histograms by a spline function. Due to the uniform normalization of all distributions, the  histograms are directly comparable. The upper row shows the distribution of the normalized temperature fluctuations within and above a Groove. We have chosen specific distances of $z/H=0.00234$ (left), $z/H=0.0048$ (middle), and $z/H=0.0099$ (right) to describe the typical dynamics of the temperature field at the half, the full and the double height of an obstacle ($h=12~\rm mm$). With respect to the typical flow situation in the Groove, it can be expected that the temperature fluctuations in this flow region behave very similar at all three distances $z/H$ compared with the smooth plate.  Solely, the distribution very close to the plate at $z/H=0.00234$ and at the lower Ra number $Ra_1=4.6 \times 10^9$ (left upper sub-figure is slightly shifted towards the temperature of the heating plate $\Theta=1$. This might be an effect of a heat transport from the sidewalls of the obstacles to the fluid flow in the Groove that distort the mean temperature field close to the roughness elements. The probability distributions of the temperature in the Notch (middle row) significantly differ from that of the smooth case. Independently, on whether or not the critical Ra number for the transition of the local flow field is exceeded, the distributions at half and full height of the obstacles are higher, narrower and clearly skewed towards the temperature of the heating plate. There is also a difference between the low and the high Ra number distribution at $z/H=0.00234$ reflecting as well the transition of the flow field within the Notch as the observed ``buckle'' in the profile of the temperature fluctuations. In the lower row, the histograms of the temperature above the top of the obstacles is shown at $z/H=0.0048$, $z/H=0.0071$, and $z/H=0.0096$. This corresponds to positions next to the top surface of an obstacle, as well as 6~mm and 12~mm above. In particular, these plots are of interest, since they show the typical distribution of the temperature fluctuations in the region, where the mean temperature profile at the higher Ra number exhibits a logarithmic trend and where they significantly deviate from the lower Ra number and from the smooth case.

Eventually, we wish to turn back to the hypothesis that roughness modifies the number and the frequency of emitted plumes at the top of the obstacles. This modification is commonly reported along with an increase of the global heat transport, however, the scaling exponent $\gamma$ remains unchanged. On the other hand, there are various experiments, in which the scaling exponent $\gamma$ changes, when a certain critical limit in Ra is exceeded. Among other experiments, such a variation in $\gamma$ has been also observed in the two equivalent RB cells in Ilmenau and Lyon \cite{Liot2016}. It seems to be clear that a laminar-turbulent transition of the boundary layer, even it appears only locally at the Top of the obstacles, will definitely affect the scaling exponent (see Fig.~\ref{Mean_top_comparison}). However, it is not confirmed that a change in the dynamics of the plume evolution does the same. In the subsequent discussion, we will focus on this we will particularly compare the plume dynamics at smooth and rough surfaces. Although, there is no final definition what a thermal plume is, people commonly consider fluid parcels that dissolve from the boundary layer at the heating (or cooling) plate and which feature a higher (or lower) temperature with respect to the surrounding fluid as such an event. Plumes arise as well at the smooth as at the rough surface and they significantly affect the heat transport in thermal convection. If they arise more frequently, and/or their typical characteristics, like their size or their energy content, change due to the roughness elements, they modify the typical bahavior of the temperature fluctuations in the fluid layer close to the plate. More precisely, one would expect an increase of the temperature fluctuations and/or a shift of their probability distribution towards the temperature at the plate surface. In this content, it might be useful to remind the profiles of the normalized standard deviation $\rm{std}(\Theta(z/H))$, in particular those measured beyond the height of the obstacles ($z/H>0.0048$). These profiles are plotted in Fig.~\ref{Tstd_lowRa} (low Ra number) and Fig.~\ref{Tstd_highRa} (high Ra number) and as we already discussed above, the temperature fluctuations are unexpectedly lower at the rough than at the smooth surface. The effect is not very large and it is more pronounced for the lower Ra number case. For the higher Ra number it almost vanishes and, in so far, there is no indication of significant higher plume activity from analyzing the standard deviation of the temperature time series. The same holds for the probability distribution of the temperature fluctuations that we plot for various distances from the plate surface in Fig.~\ref{Histograms}. The most relevant curves are the measurements beyond the height of the obstacles at $z/H=0.0099$ shown in the right sub-figures. In all three diagrams representing the distribution of the temperature fluctuations above the Groove, the Notch, and the Top region, a significant difference between the two cases with the rough plate and the case with the smooth plate appear. In summary, we can state that neither the profiles of the temperature fluctuations nor the probability distributions indicate any signature of an enhanced plume activity or any variation in their typical properties, at least for our specific roughness profile.

\begin{figure}
  \centerline{\includegraphics[width=14cm]{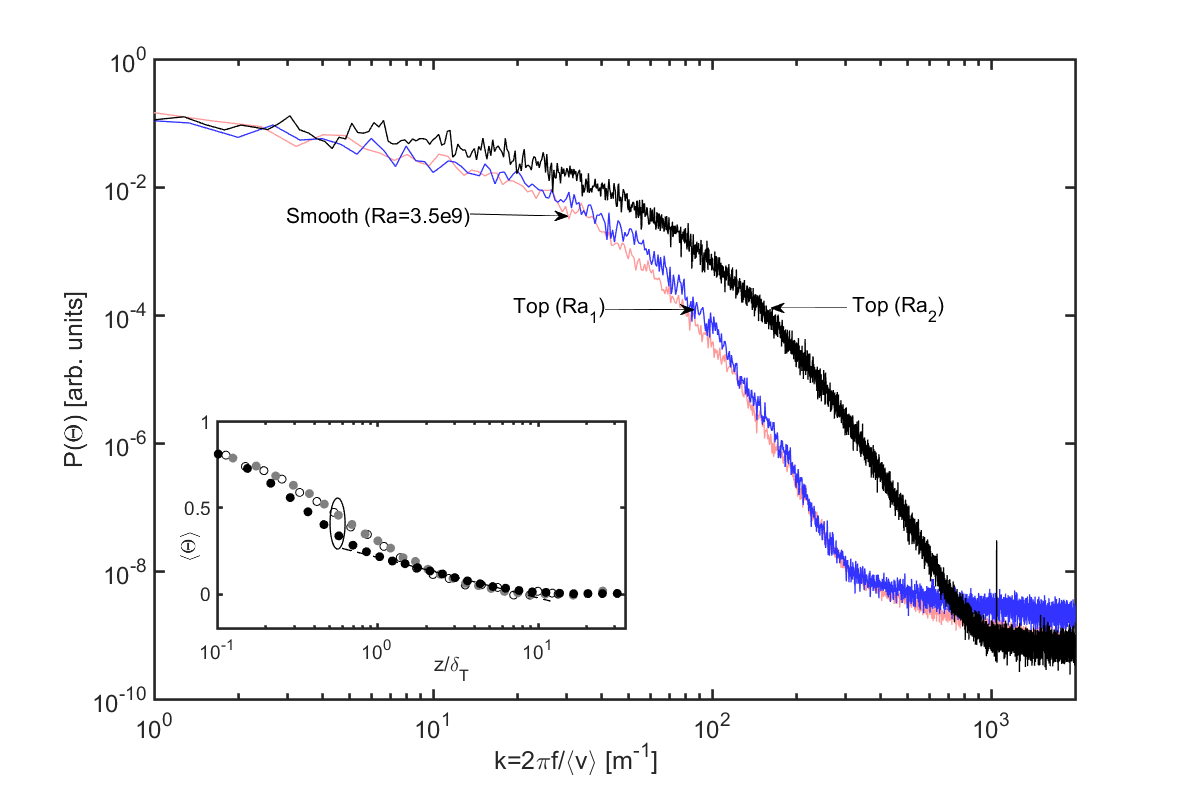}}
  \centerline{\includegraphics[width=14cm]{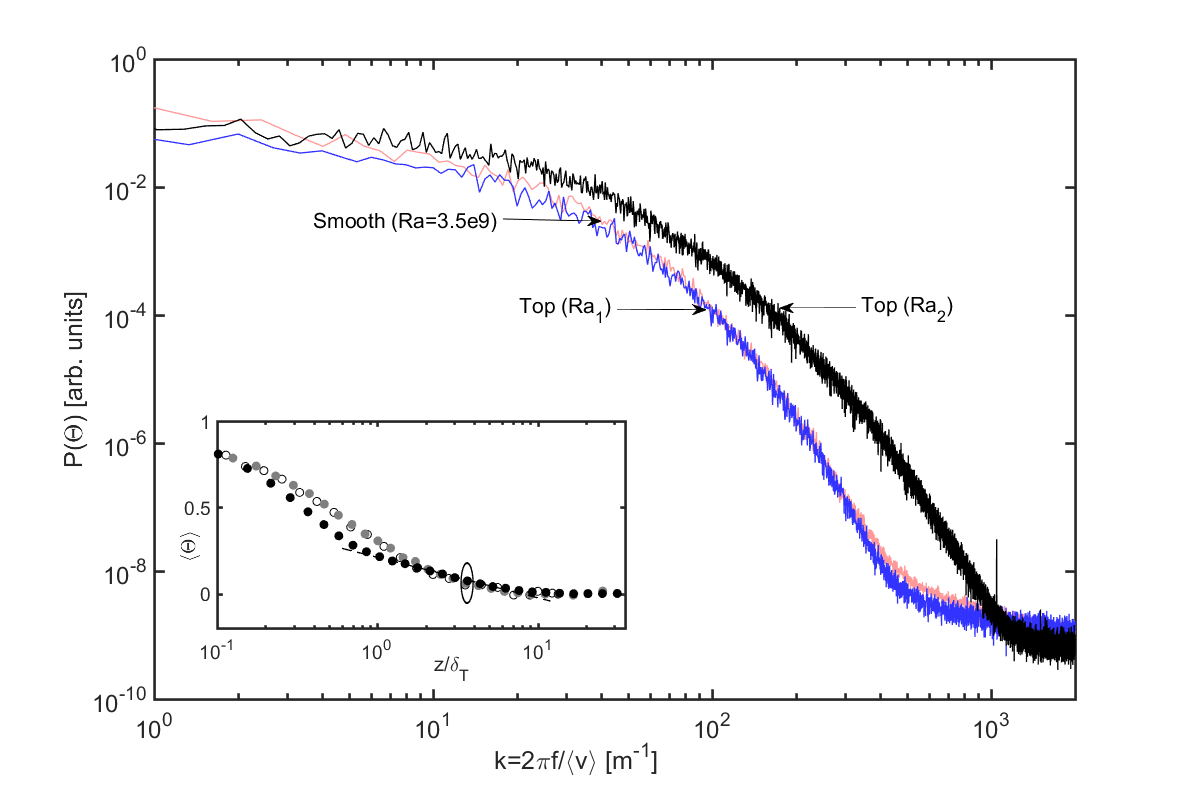}}
  \vspace{0.1cm}
  \caption{Power spectra of the temperature fluctuations at $z/\delta_T=0.6$ (upper diagramm) and $z/\delta_T=3.6$ (lower diagramm). The gray (online: blue) and the black lines represent the rough case at $Ra_1=4.6 \times 10^9$ and $Ra_2=4.7 \times 10^{10}$, respectively. For comparison, the spectra of the temperature fluctuations above a smooth surface at $Ra=3.4 \times 10^9$ has been added as light gray line (online: orange). The insets show the respective measurement positions in the mean temperature profiles as detailed shown in Fig.~\ref{Mean_top_comparison}.}
  \label{FFT}
\end{figure}

An even more sophisticated tool to study the dynamics of the temperature fluctuations in the time domain is the transformation of the signal in a Fourier space. Such an analysis provides the spectrum of the amplitude of the temperature fluctuations over the various frequencies. From this, we may associate specific properties of the spectra obtained, like its slope, its cut off or distinct peaks, with specific characteristics of the plumes and the underlying turbulent background. Since plumes are believed to preferably arise at the Top of the obstacles, we run a Fast Fourier Transformation (FFT) of selected temperature-time series above the roughness elements (TOP) for our analysis. We focus on two specific distances above the top of the obstacles. With reference to Fig.~\ref{Mean_top_comparison}, we compute a first set of FFTs at $z/\delta\approx0.6$, the normalized distance at which the profiles of the mean temperature at $Ra_1=4.6 \times 10^9$ and $Ra_2=4.7 \times 10^{10}$ obey their maximum difference. A second set of FFTs has been computed for the distance $z/\delta\approx3.6$, the position, at which the mean temperature profile at $Ra_2$ exhibits the logarithmic trend. In order to make the frequency spectra at various Rayleigh numbers comparable, we scale the frequency $f$ by the advection velocity $\langle v\rangle$ of the mean flow along the heated bottom plate and plot all spectra versus the wave number $k=2\pi f/\langle v\rangle$. The specific velocities for this normalization amounts to $\langle v_{Top,Ra1}\rangle=0.12~{\rm ms^{-1}}$, $\langle v_{Top,Ra2}\rangle=0.30~{\rm ms^{-1}}$ for the rough surface \cite{Liot2016} and $\langle v_{Smooth,Ra=3.5e11}\rangle=0.10~{\rm ms^{-1}}$ for the smooth case \cite{Li2012}. We plot the spectra at the smaller distance $z/\delta\approx0.6$ above the top of the obstacles in the upper diagram of Fig.~\ref{FFT} and the spectra at the larger distance $z/\delta\approx3.6$ in the lower diagram. For comparison, we add the spectra above a smooth surface at a slightly lower Rayleigh number $Ra=3.4 \times 10^9$. As a reminder, the insets of both diagrams show the profiles of the mean temperature for the two rough cases indicating the measurement positions that have been Fourier transformed. In both diagrams the ``rough'' and the ``smooth'' spectra at the lower Rayleigh number $Ra_1$ almost collapse. This indicates that neither the evolution and the dynamics of the plumes nor the spatial structure of the turbulent background flow is modified by the pure existence of the obstacles. Only beyond the transition, at the higher Ra number $Ra_2$, the spectra at the rough surface exhibit differ from that at the smooth one. It exhibits a more flat slope and the cut-off wave number - this is the point, at which the spectrum collapses with the noise level - is about twice as high compared with the spectra at the lower Ra number. The flatter profile and the higher cut-off wave number indicate faster and more intensive fluctuations of the temperature (and the velocity) field. The factor of ``two'' in the kick-off wave number can be interpreted as halving the typical size of the smallest turbulent structures that appear close to the rough surface. This is much more than the decrease of the boundary layer thickness between those two Ra numbers, which amounts to $\delta_T(Ra_1)=5.82~\rm{mm}$ and $\delta_T(Ra_1)=4.72~\rm{mm}$ (see Table~\ref{Gradients}), respectively, and this confirms that a real transition of the boundary layer takes place. On the other hand, we did not find any distinct bump over the entire spectra neither for the two rough nor for the smooth case. The emission of plumes happen, thus, always randomly, and this emission scheme is neither affected by introducing roughness elements nor by exceeding the critical Ra number for the boundary layer transition at the rough surface.

\section{Conclusions}

We report highly resolved temperature measurements in turbulent Rayleigh-B\'enard (RB) convection with a rough surfaces. The measurements have been undertaken in a rectangular test section at two Rayleigh  numbers $Ra_1=4.6 \times 10^9$ and $Ra_2=4.7 \times 10^{10}$. The particular idea of this work is, to study the near-wall temperature field below and beyond a critical Rayleigh number $Ra_c$, at which a transition in the scaling of the global heat transfer relation $Nu\sim Ra^{\gamma}$ has been observed as well in our experiment as in an equivalent one with water that is operated by a group at the \'Ecole Normale Sup\'erieure de Lyon \cite{Liot2016,Tisserand2011}. We used a very tiny micro-thermistor of only 130~µm in diameter and 330~µm in length for our measurements, which is about forty times smaller than the smallest thickness of the boundary layer at the top of the obstacles. The high spatial resolution of our measurements as well as the very short response time of the sensor, which are below the typical Kolmogorov microscales that occur in our experiment, enable us to reveal a potential variation of the local temperature field between the smooth and the rough cases as well as between the low and the high Ra number rough cases.

In our work, we analysed profiles of the mean temperature and the temperature fluctuations as well as the probability distribution of the temperature fluctuations at various locations with respect to the roughness elements. Our measurements demonstrate that the heat flux enhancement generally observed at rough surfaces results from an increase of the local heat transfer coefficient at the top of the obstacles. We also show that the transition of the scaling exponent $\gamma$ in the global heat flux relation $Nu\sim Ra^{\gamma}$ can be attributed to a modification of the temperature field at the top of the obstacles as well as in the flow regions in between them. Below a critical Ra number $Ra_c \approx 10^{10}$, the profile of the mean temperature $T(z)$ at the top of the obstacles basically equals with that measured at a smooth surface. Beyond this point the profile tends to exhibit a logarithmic trend, a typical signature of a turbulent boundary layer. The transition at the rough surface appears at a critical Ra number of $Ra_c \approx 10^{10}$, which is about three to four orders of magnitude lower than the limit that has been predicted for RB convection with smooth plates \cite{Grossmann2001}. The profiles of the mean temperature in the Notch and in the Groove changes as well beyond the critical Ra number. In both flow regions, the gradients are smaller and we attribute this to the onset of local convection that increases the local (and the global) heat transport. We also observed that the variation of the temperature field due to the surface roughness only covers a fluid layer with a thickness of about twice the height of the obstacles. Beyond this distance from the wall, the temperature field above the rough surface is virtually unchanged if one compare it with a smooth surface. In this far-wall region, neither the profiles of the mean temperature nor the probability distribution of the temperature fluctuations show any significant differences.

We also check the hypothesis of Du \& Tong , who discovered that the dynamics of the plume emission at a rough surface change with respect to the smooth case \cite{Du2000}. While the mean temperature field and the probability distribution of the fluctuations do not change, when roughness elements have been installed the typical size of the plumes and the frequency of their emission increases, however, even exceeding the critical Ra number for the transition of the boundary layer. For our specific roughness structure of cubic obstacles in a $l=2d$ periodical distance, we do not find any signature of a variation of the plume dynamics with respect to the introduction of roughness elements as long as one remains below the transition. This is certainly in contrast to the work, done by Du \& Tong, but might be related to the different types of roughness elements.

\bigskip

\noindent \textbf{ACKNOWLEDGMENTS}

\medskip

The authors wish to acknowledge the German Research Foundation under the grant number PU436/1-2, the German Academic Exchange Service under the project ID 57128323 and the European Union under the Grant Agreement number 312778, who gratefully supported the work reported here. We also thank the Veritas Sensors LLC providing the ultra-small micro-thermistors used in the work reported here. Moreover, we wish to thank Vigimantas Mitschunas, Sabine Abawi and Olivier Liot for their assistance to operate the experimental facility and to run the measurements.

\bigskip

\end{document}